\documentclass[preprint,12pt, authoryear]{elsarticle}

\usepackage[x11names,table,xcdraw]{xcolor}
\usepackage[
  colorlinks,
]{hyperref}

\AtBeginDocument{\hypersetup{citecolor=blue}}

\usepackage{url}
\usepackage{float}
\usepackage{hyperref}
\usepackage{amsmath}
\usepackage[graphicx]{realboxes}
\usepackage{adjustbox}
\usepackage{svg}
\usepackage{eurosym}
\usepackage[T1]{fontenc}
\usepackage{epsf}
\usepackage{caption}
\usepackage{subcaption}
\newcommand{\probP}{\text{I\kern-0.15em P}}

\usepackage{amssymb}

\begin{document}

\begin{frontmatter}



\title{Stylized Facts and Market Microstructure: An In-Depth Exploration of German Bond Futures Market}


\author[inst1,inst2]{Hamza Bodor\corref{corr1}}
\cortext[corr1]{Corresponding author.}
\ead{hamza.bodor@bnpparibas.com}
            
\author[inst2]{Laurent Carlier}
\ead{laurent.carlier@bnpparibas.com}

\affiliation[inst1]{organization={Université Paris 1 Panthéon-Sorbonne, Centre d’Economie de la Sorbonne},
            addressline={106 Boulevard de l'Hôpital}, 
            city={Paris Cedex 13},
            postcode={75642},
            country={France}}

\affiliation[inst2]{organization={BNP Paribas Corporate and Institutional Banking, Global Markets Data \& Artificial Intelligence Lab},
            addressline={20 boulevard des Italiens}, 
            city={Paris},
            postcode={75009},
            country={France}}

\begin{abstract}
This paper presents an in-depth analysis of stylized facts in the context of futures on German bonds. The study examines four futures contracts on German bonds: Schatz, Bobl, Bund and Buxl, using tick-by-tick limit order book datasets. It uncovers a range of stylized facts and empirical observations, including the distribution of order sizes, patterns of order flow, and inter-arrival times of orders. The findings reveal both commonalities and unique characteristics across the different futures, thereby enriching our understanding of these markets. Furthermore, the paper introduces insightful realism metrics that can be used to benchmark market simulators. The study contributes to the literature on financial stylized facts by extending empirical observations to this class of assets, which has been relatively underexplored in existing research. This work provides valuable guidance for the development of more accurate and realistic market simulators.
\end{abstract}



\end{frontmatter}


\newpage
\section{Introduction}
For decades, the study of stylized facts in financial markets has been a primary focus for researchers and practitioners \citep{ guillaume1997bird,cont2001empirical, bouchaud2002statistical}. Stylized facts refer to the common statistical properties and patterns observed across various financial markets, such as fat-tailed distribution of returns, volatility clustering, and correlation of asset returns \citep{mantegna1999introduction, chakraborti2011econophysics}. These properties serve as essential building blocks for the development of realistic and robust financial models, which in turn facilitate better understanding and management of market risks.

Realism metrics play a crucial role in assessing the validity of market models and their ability to reproduce stylized facts \citep{vyetrenko2020get}. Their importance is twofold: (1) they enable researchers to gauge the performance of their models against real-world market data, ensuring that the models capture the underlying market dynamics accurately \citep{lux1998scaling, harmon2011predicting, abergel2016limit}; and (2) they provide a benchmark for comparing and improving existing models, which can lead to more reliable forecasts and better-informed decision-making processes \citep{lebaron2006agent, brandouy2012reexamination}.
However, most existing studies on stylized facts and realism metrics tend to focus on specific markets, such as foreign exchange (FX) \citep{guillaume1997bird, engle1997forecasting, ballocchi1999intraday, aloud2013stylized} and equity markets \citep{bouchaud2002statistical, potters2005financial, chakraborti2011econophysics, bouchaud2018trades}.

In the following sections, we will explore the stylized facts that are particularly relevant for futures on German bonds, a distinct market with its own characteristics. For instance, these futures are known for their high liquidity, their sensitivity to changes in interest rates, and their role as a benchmark for European fixed income securities. By extending the analysis to this market, we aim to contribute to the growing body of knowledge on financial stylized facts and further enhance the development of realistic market models that can accommodate a broader range of financial instruments and market conditions.

\subsection{Overview of Futures on Treasury Bonds}

Bonds are a type of debt security that represents a contractual obligation between a borrower and a lender. They are issued by corporations, governments, and other entities to raise capital for various purposes, such as financing projects, expanding operations, or refinancing existing debt. Bonds typically have a fixed maturity date, at which point the issuer is obligated to repay the principal amount borrowed to the bondholder. In addition, bonds pay periodic interest payments to the bondholder, which are typically fixed at the time of issuance and based on a percentage of the bond's face value. Bonds are traded in financial markets and are subject to fluctuations in price and yield based on changes in interest rates, credit ratings, and other factors \citep{labuszewski2013understanding}.

Futures are a type of financial derivatives that allow investors to buy or sell an underlying asset at a predetermined price and date in the future \citep{hull1993options}. Futures contracts are standardized agreements that specify the quantity, characteristics, and delivery date of the underlying asset, as well as the price at which the transaction will occur. Futures are used by investors to hedge against price fluctuations in the underlying asset, as well as to speculate on the future price movements of the asset. 

In the context of bonds, futures contracts are used to trade fixed income securities, such as government bonds, at a future date and price. Bond futures allow investors to take a position on the future price of the underlying bond, without actually owning the bond itself. This can be useful for investors who want to hedge against interest rate risk or speculate on changes in bond prices. Bond futures are traded on exchanges, such as the Chicago Mercantile Exchange (CME) and the Eurex Exchange, and are sensitive to fluctuations in interest rates, credit ratings, or other market factors, mirroring the underlying bond's behavior.

Bond futures can be used by investors to manage risk, enhance returns, and diversify their portfolios. In fact, within the specific framework of bond futures, an investor with substantial bond holdings, who anticipates potential interest rate hikes (which would precipitate a decrease in their bond prices), may opt to sell or \emph{short} bond futures contracts. In the event of an interest rate increase, the value of the investor's bond portfolio would diminish, but the short futures position would appreciate in value, thereby neutralizing the loss from the bond portfolio.
On the other hand, an investor intending to purchase bonds in the future but apprehensive about potential interest rate declines (which would increase the price of bonds) may choose to purchase or \emph{go long} bond futures contracts. If interest rates indeed fall, the investor would have to incur a higher cost for the bonds, but the long futures position would appreciate in value, counterbalancing the increased cost of the bonds.

One has to differentiate futures from forward contracts on bonds by considering their unique characteristics and trading mechanisms. Forward contracts are privately negotiated, customizable, and traded over-the-counter or off-exchange, with the payment fully dependent on the counterparty, introducing a higher rate of default risk. Conversely, futures contracts are exchange-traded, standardized agreements with specified contract units, expiration, tick sizes, and notional values. They are actively traded, regulated, and devoid of counterparty risk due to the payment guarantee by the exchange clearing house. Another differentiating feature of futures contracts is the concept of margin and margin calls. Margin is the amount of money or collateral that a trader must deposit with a broker or exchange to cover potential losses from their trading activities. Margin requirements are set by the exchange or broker and vary depending on the type of asset being traded, the volatility of the market, and the trader's experience and risk appetite. Margin calls occur when the value of a trader's account falls below the required margin level, and the broker or exchange demands additional funds to cover the potential losses. These features, along with the ease of entering and exiting positions, have made futures contracts an integral component of the global economy, attracting a large number of market participants. In fact, Table \ref{tab: most active products eurex 0723} enumerates the ten most active products on the Eurex exchange, including three futures contracts (highlighted in green) on German bonds. These products are ranked closely behind EURO STOXX 50 Index derivatives, one of the most traded products worldwide. Further details about the specifications of these futures contracts will be explored in subsequent sections. 

\begin{table}[htbp]
\center
\small
\begin{tabular}{llll}
\rowcolor[HTML]{FFFFFF} 
\hline
\multicolumn{1}{c}{\cellcolor[HTML]{FFFFFF}{\color[HTML]{000000} \textbf{\begin{tabular}[c]{@{}c@{}}Product\\ Name\end{tabular}}}} & \multicolumn{1}{c}{\cellcolor[HTML]{FFFFFF}{\color[HTML]{000000} \textbf{\begin{tabular}[c]{@{}c@{}}Product \\ Type\end{tabular}}}} & \multicolumn{1}{c}{\cellcolor[HTML]{FFFFFF}{\color[HTML]{000000} \textbf{\begin{tabular}[c]{@{}c@{}}Product \\ ID\end{tabular}}}} & \multicolumn{1}{c}{\cellcolor[HTML]{FFFFFF}{\color[HTML]{000000} \textbf{\begin{tabular}[c]{@{}c@{}}\# Traded \\ Contracts\end{tabular}}}} \\ \hline
\begin{tabular}[c]{@{}l@{}}EURO STOXX 50 \\Index Options\end{tabular}                                                              & Option                                                                                                                              & OESX                                                                                                                              & 18,847,107                                                                                                                                 \\ \hline
\begin{tabular}[c]{@{}l@{}}EURO STOXX 50 \\Index Futures\end{tabular}                                                              & Future                                                                                                                              & FESX                                                                                                                              & 15,358,844                                                                                                                                 \\ \hline
\rowcolor[HTML]{B6F1A6} 
Euro-Bund Futures                                                                                                                  & Future                                                                                                                              & FGBL                                                                                                                              & 14,532,480                                                                                                                                 \\ \hline
\rowcolor[HTML]{B6F1A6} 
Euro-Bobl Futures                                                                                                                  & Future                                                                                                                              & FGBM                                                                                                                              & 11,342,587                                                                                                                                 \\ \hline
\rowcolor[HTML]{B6F1A6} 
Euro-Schatz Futures                                                                                                                & Future                                                                                                                              & FGBS                                                                                                                              & 8,900,982                                                                                                                                  \\ \hline
Euro-OAT-Futures                                                                                                                   & Future                                                                                                                              & FOAT                                                                                                                              & 3,342,300                                                                                                                                  \\ \hline
EURO STOXX Banks                                                                                                                  & Future                                                                                                                              & FESB                                                                                                                              & 2,961,332                                                                                                                                  \\ \hline
Euro-BTP Futures                                                                                                                   & Future                                                                                                                              & FBTP                                                                                                                              & 2,921,050                                                                                                                                  \\ \hline
\begin{tabular}[c]{@{}l@{}}EURO STOXX Banks\\ Options\end{tabular}                                                                & Option                                                                                                                              & OESB                                                                                                                              & 2,520,810                                                                                                                                  \\ \hline
\begin{tabular}[c]{@{}l@{}}Options on Euro-Bund\\ Futures\end{tabular}                                                             & Option                                                                                                                              & OGBL                                                                                                                              & 2,409,951                                                                                                                                  \\ \hline
\end{tabular}
\caption{Most active products on Eurex exchange during July 2023}
\label{tab: most active products eurex 0723}
\end{table}

When trading bond futures, investors should be aware of two important concepts: open interest and volume. Open interest refers to the total number of outstanding futures contracts that have not been closed out or delivered. It is a measure of the market's overall interest in a particular futures contract and can be used to gauge the level of liquidity and trading activity in the market. Volume, on the other hand, refers to the total number of contracts that have been traded during a given period, such as a day or a week. Volume is a measure of the level of trading activity in the market and can be used to identify trends and patterns in the market. High levels of open interest and volume can indicate a liquid and active market, which can make it easier for investors to buy or sell futures contracts at a fair price. Conversely, low levels of open interest and volume can indicate a less liquid market, which can make it more difficult for investors to trade futures contracts and may result in wider bid-ask spreads and higher transaction costs. By monitoring open interest and volume, investors can gain valuable insights into the market's overall health and make more informed trading decisions.

The delivery of futures contracts can be settled in one of two ways: physical delivery or cash settlement. Physical delivery involves the actual transfer of the underlying asset from the seller to the buyer at the expiration of the contract. In the case of bond futures, this would involve the delivery of a bond with the attributes specified in the contract. The seller would be responsible for delivering the bond to the buyer, while the buyer would be responsible for paying the delivery price. Cash settlement, on the other hand, involves the payment of a cash amount equal to the difference between the contract price and the market price of the underlying asset at the expiration of the contract. This method is often used when the underlying asset is difficult to deliver or when the cost of delivery is prohibitively high. 

One particularity of the futures on bonds is that the short position (the party that sells the futures contract), in the case of physical delivery, has a choice among a basket of bonds for delivery to the long position (the party that buys it). The range of possible bonds to deliver depends on the remaining lifetime of the bond, and is variable for each futures contract, for examples Table \ref{tab:contracts specification} gives the range of deliverable maturities for each future instrument on German bonds.

These deliverable bonds, despite originating from the same issuer, are not identical due to variations in their coupons, maturities, and consequently, their prices. \textit{The conversion factor} plays a crucial role at the time of delivery, as it is used to compute the final delivery price. Essentially, the conversion factor establishes a hypothetical trading price for a bond, assuming its yield were either six or four percent on the delivery day (depending on the contract).

One of the fundamental assumptions inherent in the conversion factor formula is the expectation of a flat yield curve on the delivery date. Additionally, this flat yield curve is presumed to precisely align with the notional coupon of the futures contract. Under these assumptions, all bonds within the delivery basket would theoretically hold equal appeal for delivery. However, in practice, the yield curve rarely exhibits a flat profile, leading to an unintended bias introduced by the conversion factor that favors certain bonds for delivery over others. Consequently, this bias grants the short position the liberty to select, from the pool of deliverable bonds, those that yield the most advantageous outcomes. This concept is referred to as the \textit{Cheapest-to-Deliver (CTD)} and reflects the short position's optimization of self-interest in minimizing their net cost in a bond futures contract. The CTD represents the bond among the available deliverable bonds that has the lowest net cost from the point of view of the contract's short.

\subsection{Assets Description}
\label{sec:SFs_assets_desc}

The assets explored in this study comprise futures contracts on German bonds, issued by the Federal Republic of Germany and traded on the Eurex exchange, specifically including Schatz, Bobl, Bund, and Buxl. All futures contracts described carry a face value of 100,000\euro~ and are based on German debt instruments with varying residual maturities and coupon rates. \textbf{Schatz Future}, representing short-duration contracts, have underlying securities with a residual maturity of 1.75 to 2.25 years and a standardized 6\% coupon rate.\footnote{It is pertinent to note that Germany has not issued debt securities with a coupon rate as elevated as 6\% in recent times. Therefore, to reconcile this discrepancy, the contract values are methodically adjusted to reflect the prevailing market price of a bond with a 6\% coupon, utilizing the conversion factor as referenced earlier.} \textbf{Bobl Future} represent medium-term contracts with 4.5 to 5.5 years of remaining tenure and also feature a 6\% coupon rate. Moving towards longer durations, \textbf{Bund Future} are long-term contracts based on securities that span 8.5 to 10.5 years of residual maturity and maintain the same 6\% coupon rate. Lastly, \textbf{Buxl Future} are associated with an even lengthier period of 24.0 to 35.0 years and utilize a distinctive 4\% coupon rate.

\begin{figure}[htbp]
\includegraphics[clip, trim=3.0cm 0.5cm 1.5cm 10.0cm,width=\textwidth]{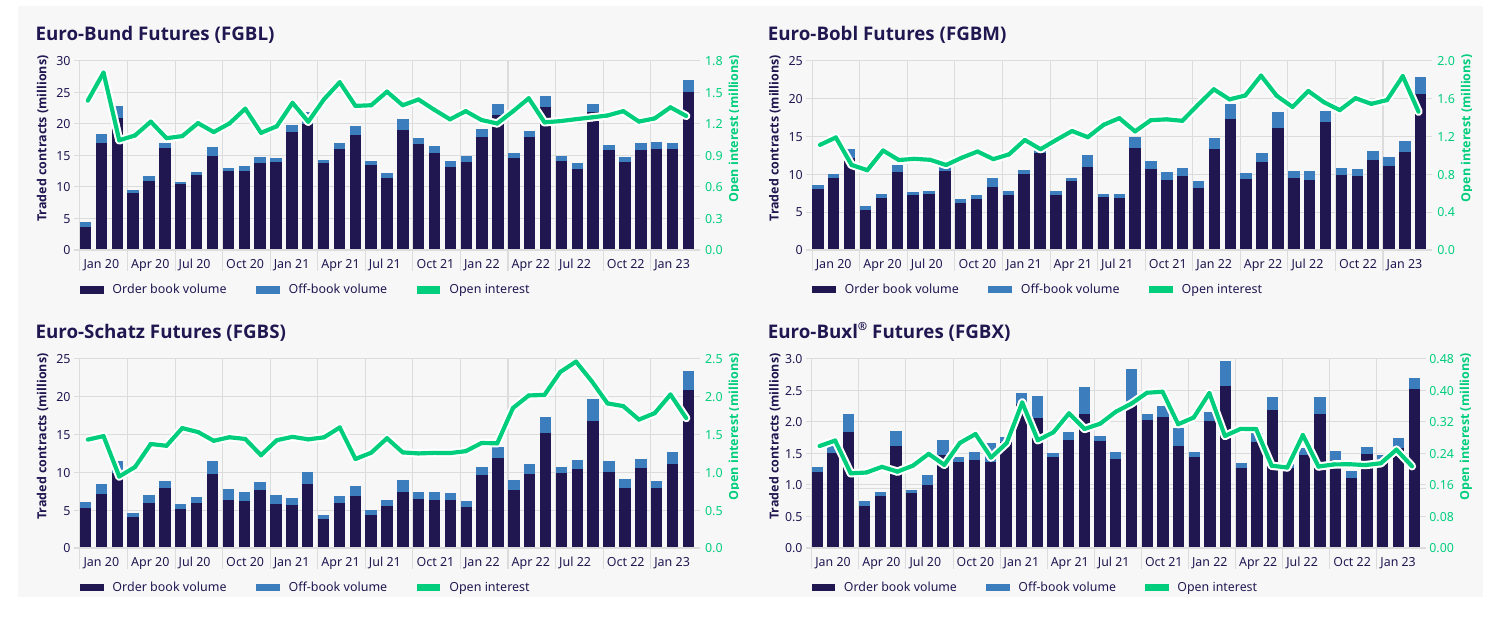}
\caption{Traded contracts and open interest evolution of the four futures on German bonds. }
\label{fig: compartison liquidity}
\end{figure}

Figure \ref{fig: compartison liquidity} shows the evolution of open interest and volumes traded of each of the aforementioned futures contract on the last two years, where we can notice an overall stability of the open interest and volumes, with some spikes in the months of March, June, September and December\footnote{Figure extracted from Eurex facts sheet on fixed income derivatives \citep{Eurex_benchmark}, 2023 edition.}. These months correspond to the possible delivery periods for these futures, and are thus known to exhibit heightened activity. It is also noteworthy that the Bund and Buxl futures are the most and least active ones, respectively.

\section{Data Description}

For the purpose of this study, four futures contracts on German bonds were utilized. The data collection process spanned active trading days throughout the year of 2021, encompassing trading sessions that took place during the most active hours of the Eurex exchange, namely from 9 to 18 (CET).  \\Table \ref{tab:contracts specification} presents several characteristics of the data collected and specifications of each contract.

\begin{table}[htbp]
\setlength{\tabcolsep}{4pt}
    \centering
    \begin{adjustbox}{center,%
      width=0.9\linewidth}
    \small 
    \begin{tabular}{lccccccc}
        \hline
        \multicolumn{1}{c}{\textbf{\begin{tabular}[c]{@{}c@{}}Product\\ Name\end{tabular}}} & \multicolumn{1}{c}{\textbf{\begin{tabular}[c]{@{}c@{}}Product \\ ID\end{tabular}}} & \multicolumn{1}{c}{\textbf{\begin{tabular}[c]{@{}c@{}}Deliverable  \\ Maturities \\(Years)\end{tabular}}} & \multicolumn{1}{c}{\textbf{\begin{tabular}[c]{@{}c@{}}Tick \\ Size\end{tabular}}} & \textbf{\begin{tabular}[c]{@{}l@{}}Average\\ \# Events \\per day\end{tabular}} & \textbf{\begin{tabular}[c]{@{}l@{}}Average\\ \# Trades \\per day\end{tabular}} & \textbf{\begin{tabular}[c]{@{}l@{}}Average\\ Spread \\(Ticks)\end{tabular}} & \textbf{\begin{tabular}[c]{@{}c@{}}Average\\ Event \\ Size \end{tabular}} \\ \hline
        \multicolumn{1}{c}{\begin{tabular}[c]{@{}c@{}}Euro-Schatz \\ Futures\end{tabular}}                                                                    & FGBS                & 1.75 to 2.25                                                                                                                                                                              & 0.005        & $1.29 \times 10^5$ & $5.56 \times 10^3$ & 1.004 & 88.9 \\ \hline
        \multicolumn{1}{c}{\begin{tabular}[c]{@{}c@{}}Euro-Bobl \\ Futures\end{tabular}}                                                                       & FGBM                & 4.5 to 5.5                                                                                                                                                                                & 0.01     & $5.84 \times 10^5$ & $1.03 \times 10^4$ & 1.005 & 21.2 \\ \hline
        \multicolumn{1}{c}{\begin{tabular}[c]{@{}c@{}}Euro-Bund \\ Futures\end{tabular}}                                                                         & FGBL                & 8.5 to 10.5                                                                                                                                                                              & 0.01     & $1.86 \times 10^6$ & $3.90 \times 10^4$ & 1.018 & 5.78 \\ \hline
        \multicolumn{1}{c}{\begin{tabular}[c]{@{}c@{}}Euro-Buxl \\ Futures\end{tabular}}                                                                     & FGBX                & 24.0 to 35.0                                                                                                                                                                            & 0.02     & $9.78 \times 10^5$ & $2.06 \times 10^4$ & 1.355 & 1.38 \\ \hline
    \end{tabular}
    \end{adjustbox}
    \caption{Characteristics of the four futures subject to our study.}
    \label{tab:contracts specification}
\end{table}

Furthermore, trades were limited to \textit{on-book} trades, as these are the ones reflected in the order book. All other types of trades, mainly \textit{off-book} trades\footnote{Refer to transactions that are not executed on the centralized exchange platform. Off-book trades can occur over-the-counter (OTC) or through alternative trading systems (ATS) where parties negotiate and execute the trade directly. They are typically not visible on the exchange's order book and are often used for large or specialized transactions.}, have been omitted from this study.

\subsection{Order Flow Construction}

The data at our disposal consist solely of updates of the limit order book, up to the fifth limit, and a daily trade history. To extract the precise sequence of order flow events, we engaged in a manual process, differentiating each update as either a provision of liquidity, limit orders, or consumption of liquidity. The latter category was further divided into cancel and market orders.

\begin{figure}[htbp]
\centering
\includegraphics[angle=90,origin=c, height = 0.78 \textheight]{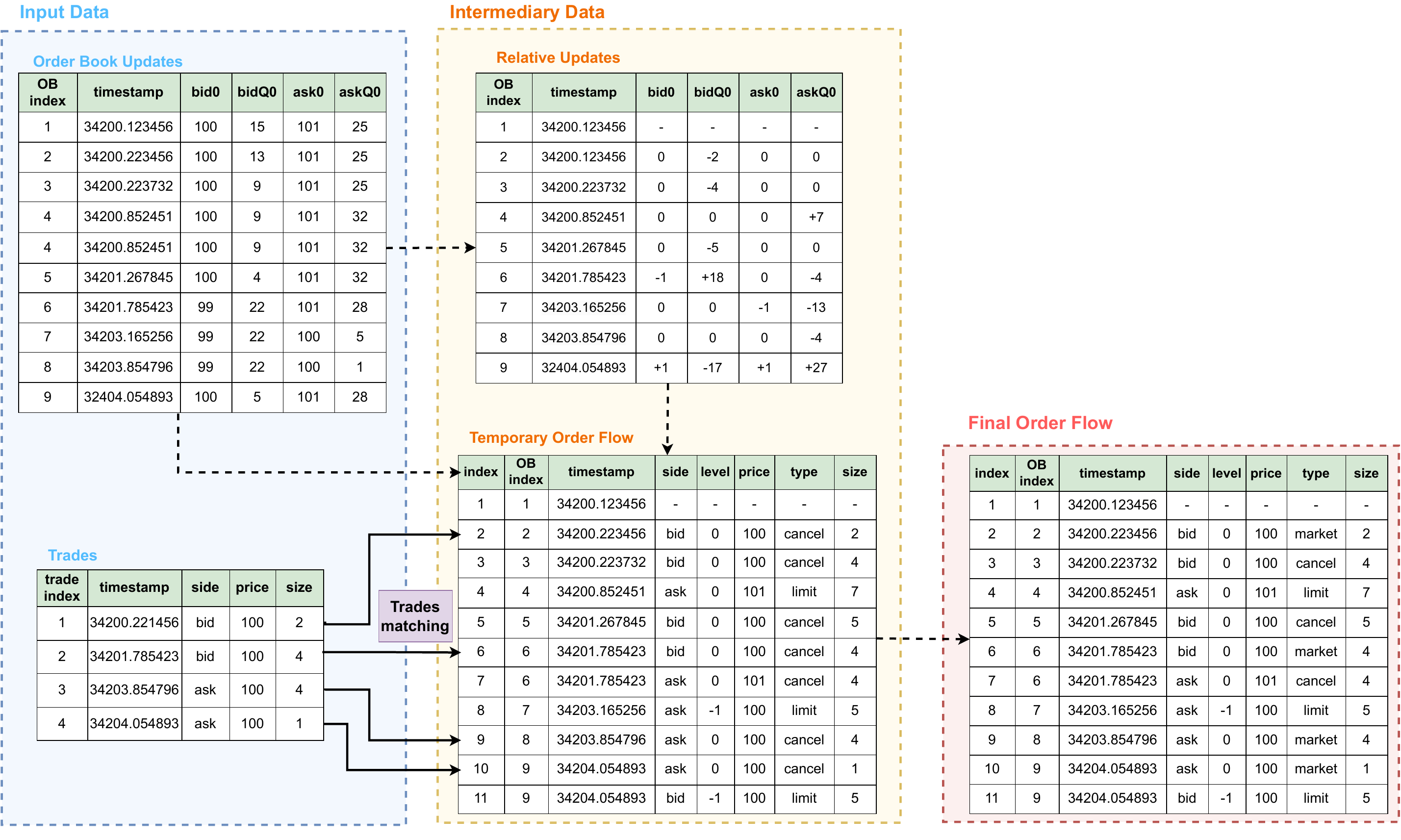}
\caption{Steps of order flow extraction}
\label{fig:steps of order flow}
\end{figure}

Figure \ref{fig:steps of order flow} delineates the procedure for constructing order flow from the order book updates (wherein, for simplicity, only the best ask and bid prices are presented) and trades.\footnote{Data recording all the transactions that took place in the considered period.} This commences with the acquisition of \emph{Relative updates} data, encompassing the differences in quantities and prices at each limit and for each update. Five discernible scenarios may arise (between brackets index of updates from \emph{Order Book Updates} data in Figure \ref{fig:steps of order flow} to illustrate each scenario when possible):

\begin{itemize}
\item A single quantity update without any price update (updates 2 or 4 for example), corresponding to a mere increment or decrement of the available quantity on a given level. This is the most rudimentary and common case, labeled as \emph{limit} if the difference is positive and \emph{cancel} otherwise.
\item Dual quantity updates on the same side with opposite values,  indicating market participants going deeper or shallower in the order book's depth. This is typically rendered as two updates: cancellation of the original order and the reinstatement of a new limit order at the revised price. However, the Eurex exchange allows this to be accomplished through a \emph{stop-limit order}.\footnote{Further details on the order types available on the Eurex exchange can be accessed at \url{https://www.eurex.com/ex-en/trade/order-book-trading/order-types}.} In our labeling process, this kind of event will be denoted in two segments: \emph{cancel} order followed by a \emph{limit} event.
\item Stop orders may also be executed as \emph{stop market orders}. Unlike the preceding scenario, which does not affect the overall liquidity on the book, these orders (order book update 6 for example) consume liquidity in two ways. Initially, they impact the liquidity on one side by canceling it, and subsequently, they affect the opposite side, where the identical quantity consumes liquidity from the best price, culminating in a trade. This process is categorized into two events: first, as a \emph{cancel}, followed by a \emph{market} order, with the latter being treated as a \emph{cancel} until the subsequent phase.

\item Updates combining multiple quantity changes with price updates, generally occurring when the mid-price shifts. These updates, comprising around 1\% of total updates, are further delineated by considering liquidity availability by price rather than by depth level. This captures all liquidity updates, especially those incited by stop market orders altering the mid-price. Three sub-cases are distinguished:

\begin{enumerate}
\item Only prices of one side are updated (order book update index 7), leading to a change in the mid-price. This indicates that the best price on the changed side has been altered, either causing a spread increase in the case of a \emph{cancel/market} order or a spread decrease with a new limit order within the spread. In the latter instance, it is labeled with a negative value of \emph{level}, corresponding to the difference in spreads from the last best price. For instance, if the best ask price is 100 and the new best ask price is 98, this event will be labeled with a \emph{level} of -2.
\item Only prices of one side change without a mid-price shift, corresponding to the emergence or disappearance of a deeper price. Here, identification of the price that has appeared or disappeared is the only requirement.
\item All prices are updated (order book update 9), generally when an aggressive limit on one side comes in at a price lower than the best price of the opposing side, resulting in a trade transaction that fully consumes the best opposite side. This leads to a spread increase, paired with a limit order within the spread for the remaining quantity of the original order, pushing the spread back to its previous value.
\end{enumerate}
\end{itemize}

Afterward, we obtain the \emph{Temporary Order Flow} data, consisting of a list of events that are either limit orders or cancel orders. For the latter, we identify the events that correspond to market orders by matching each trade in the \emph{Trades data} with all the cancel events that share the same size, side, and price in a window of size $10ms$. The closest event (in terms of time difference) is then labeled as a market order (We could match around 99\% of trades using this method\footnote{It is pertinent to note that if a significantly lower matching rate was encountered, alternative trade matching methodologies might be adopted. For instance, employing an aggregation strategy of multiple successive trades could be considered. A comprehensive study on varied order flow reconstruction methodologies can be found in \citet{toke2016reconstruction}.}). This allows us to obtain the \emph{Final Order Flow}.

To test the accuracy of the labeling method, we have built a market replayer that starts from a given order book state and replays the sequence of events in the order flow thanks to a matching engine that reflects the order into the order book. We consistently achieve a 100\% match between the replayed order book and the historical one, thereby confirming the accuracy of the order flow construction method.

\subsection{Time Types}

In the analysis of limit order book data, it is crucial to understand the different types of time scales that are often used. These time scales provide different perspectives on the dynamics of the market and can reveal different aspects of market behavior. Here, we introduce four commonly used time scales:

\begin{itemize}	
	\item \textbf{Calendar Time:} This is the most straightforward time scale and refers to the actual chronological time. It is continuous and takes into account all periods, including times when the market is inactive, such as overnight or during weekends.
	
	\item \textbf{Event Time:} This scale is based on the sequence of events in the market. Each event, such as the placement or cancellation of an order, increments the event time by one. This time scale is particularly useful for studying the microstructure of the market, as it directly reflects the activity of market participants.
	
	\item \textbf{Tick Time:} Similar to event time, but it only increments when there is a change in the best bid or ask price. This time scale is useful for studying price changes and volatility, as it filters out events that do not impact the price.
	
	\item \textbf{Trade Time:} It increments each time a trade is executed. This time scale focuses on the actual transactions in the market, ignoring other types of events such as order placements or cancellations that do not result in trades. 
\end{itemize}

Unless explicitly stated otherwise, the term \textit{time} in the following sections refers to calendar time.

\subsection{Terminology and Symbols}

Within the scope of this paper, we delineate specific notations and terminology to enhance comprehension. Let $t$ represent any given time point. We designate $b_t$ as the best bid price and $a_t$ as the best ask price at time $t$. The mid-price is then represented by $m_t = \frac{a_t + b_t}{2}$. Choosing a time scale $\Delta t$, the log return can be expressed as $R_{t, \Delta t} = \ln m_{t+\Delta t} - \ln m_t$. Return volatility, denoted by $\sigma_{\tau, \Delta t}$, corresponds to the standard deviation of these returns over a time period $\tau$. 

We also introduce notation for the available volumes at the best bid and ask prices, symbolized as $V_b$ and $V_a$ respectively. $V_\tau$ is defined as the cumulative traded volumes over a time period $\tau$. Lastly, the notation $p(.)$ is employed to represent the probability density function of an observed quantity.

\section{Stylized Facts About Asset Return Distributions}\label{stylized-facts-about-asset-return-distributions}

\subsection{Absence of Autocorrelation}

The absence of autocorrelation is a stylized fact that refers to the observation that there is no correlation between time series data and its lagged versions. 
Mathematically, the absence of autocorrelation can be expressed as the autocorrelation being equal to zero at all lags:

\begin{equation*}
\rho_{k, \Delta t} = \frac{\mathrm{Cov}(R_{t, \Delta t}, R_{t-k\Delta t, \Delta t})}{\mathrm{Var}(R_{t, \Delta t})} = 0
\end{equation*}

\begin{figure}[H]
	\centering
	\includegraphics[width=0.8\textwidth]{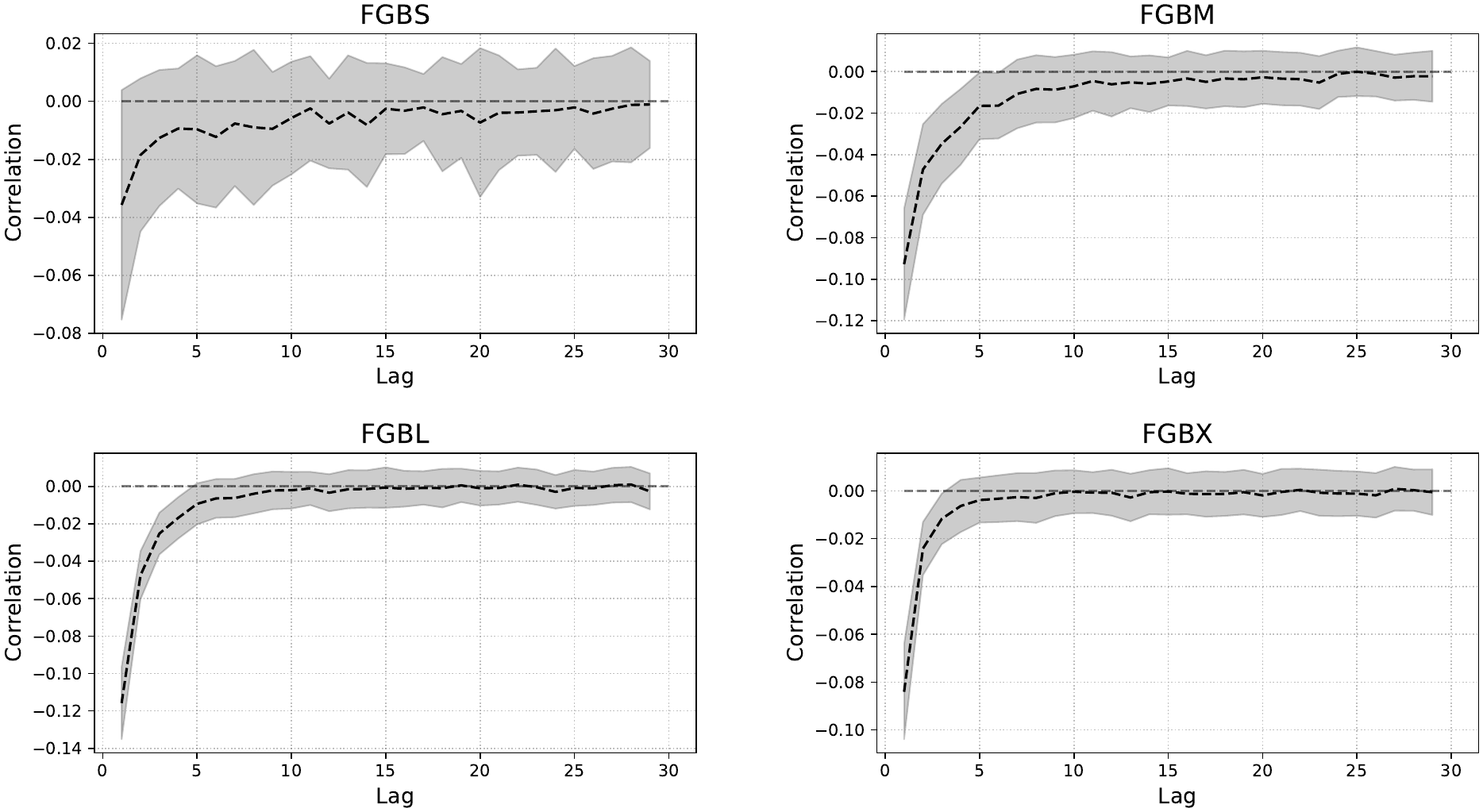}
	\caption{Autocorrelation function of tick-by-tick returns.}
	\label{fig:Autocorrelation of returns - mid - tbt}
\end{figure}

Figure \ref{fig:Autocorrelation of returns - mid - tbt} shows the curve of autocorrelation values between tick-by-tick returns (returns computed between tick-by-tick prices, without sampling) alongside with the  95\% confidence interval. It shows that the values of autocorrelation are very close to zero, which proves the stylized fact of the absence of autocorrelation between returns, except for the first lag, which shows a considerable negative value. The same behavior is noticed in \citet{abergel2016limit} on equity markets, and is due to the phenomenon of bid-ask bounce.

\begin{figure}[htbp]
	\centering
	\includegraphics[width=0.9\textwidth]{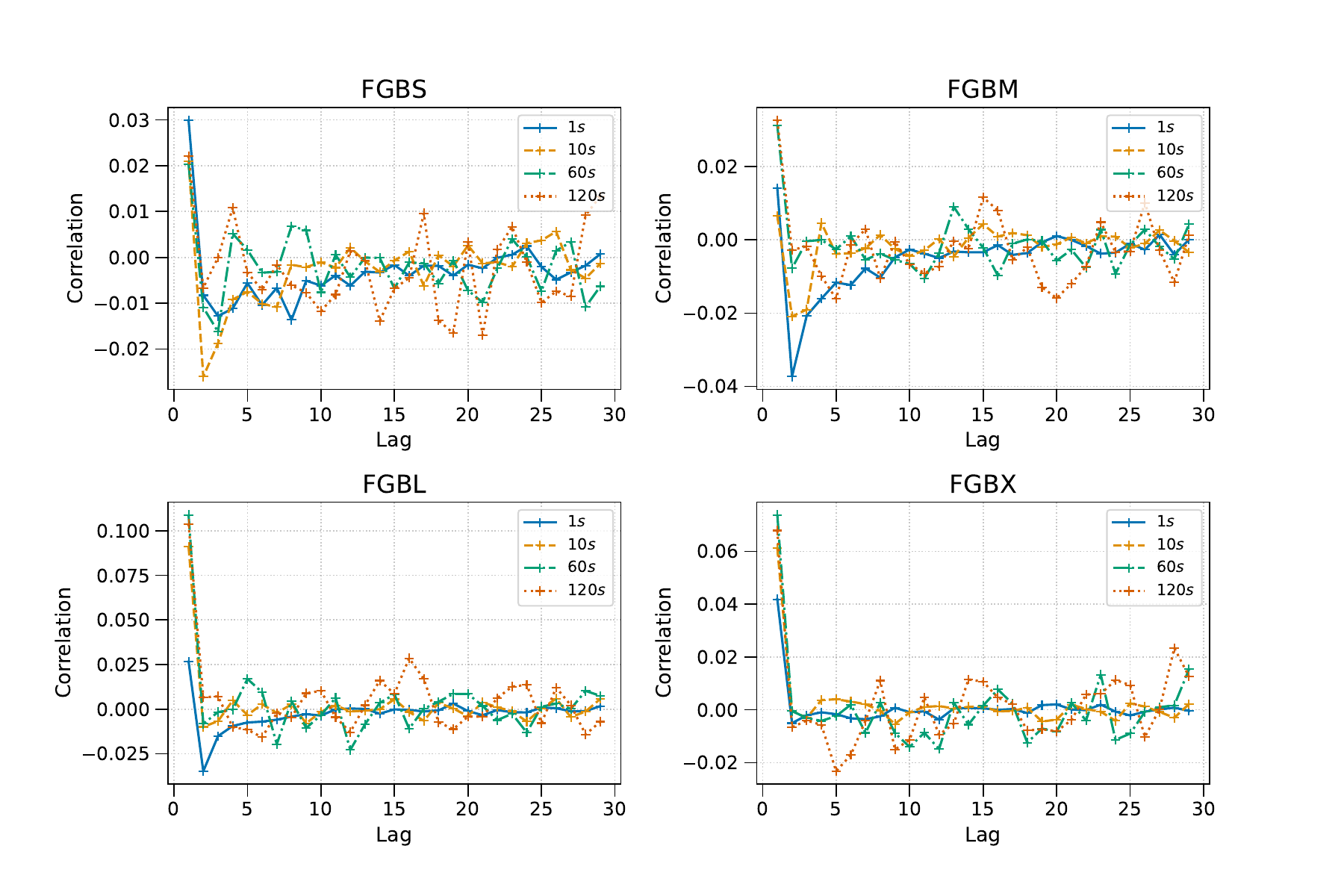}
	\caption{Autocorrelation function of sampled returns for different sampling frequencies.}
	\label{fig:Autocorrelation of returns - mid}
\end{figure}

Figure \ref{fig:Autocorrelation of returns - mid} demonstrates the absence of autocorrelation between returns across all lag values and for various sampling frequencies $\Delta_t$. It further reveals that the autocorrelation coefficients are very close to zero for all time lags. The near-zero values across all lag times indicate efficient market behavior with no predictable patterns in price movements.

These two figures display a complete absence of  autocorrelation between realized returns. This stylized fact is crucial, as it ensures that there are no basic opportunities for arbitrage, which could otherwise lead to unrealistic market simulators, where training on such simulators may lead to learning some patterns that are only valid on the simulator.

\subsection{Positive Correlation between Volume and Volatility}

The positive correlation between volume and volatility stylized fact refers to the observed tendency of trading volume and asset price volatility to be positively related. In other words, periods of high trading activity often coincide with elevated levels of price volatility \citep{brandouy2012re}.

\begin{figure}[htbp]
	\centering
	\includegraphics[width=0.8\textwidth]{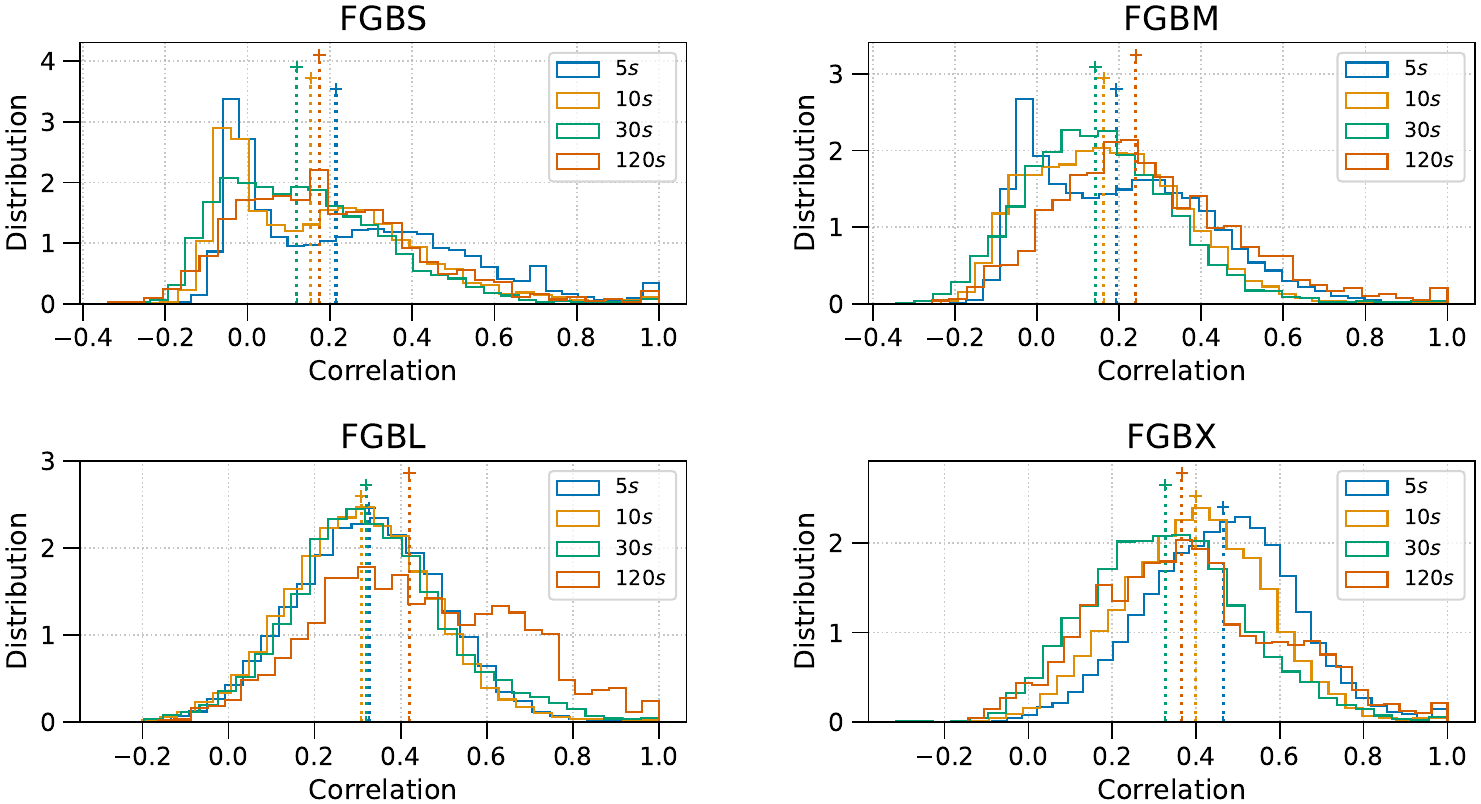}
	\caption{Distribution of correlations between volumes and volatility on windows of size 10 minutes.}
	\label{fig:Volume-Volatitlity Positive Correlation- Histogram of correlations}
\end{figure}

Figure \ref{fig:Volume-Volatitlity Positive Correlation- Histogram of correlations} presents a histogram depicting correlations between trading volume and price volatility at various sampling frequencies. Data is subdivided into equal periods, each with a size of $\tau = 100$ times the sampling frequency, with the correlation calculated, within each period, between $\sigma_{\tau, \Delta}$ (volatility) and $V_{\tau}$ (volume sum in period $\tau$). Dotted points represent median values of distributions, revealing a skew towards positive correlation values across varying time scales, thus affirming the established stylized fact for all instruments. FGBL and FGBX exhibit a more pronounced correlation.

\begin{figure}[htbp]
\centering
\includegraphics[width=0.66\textwidth]{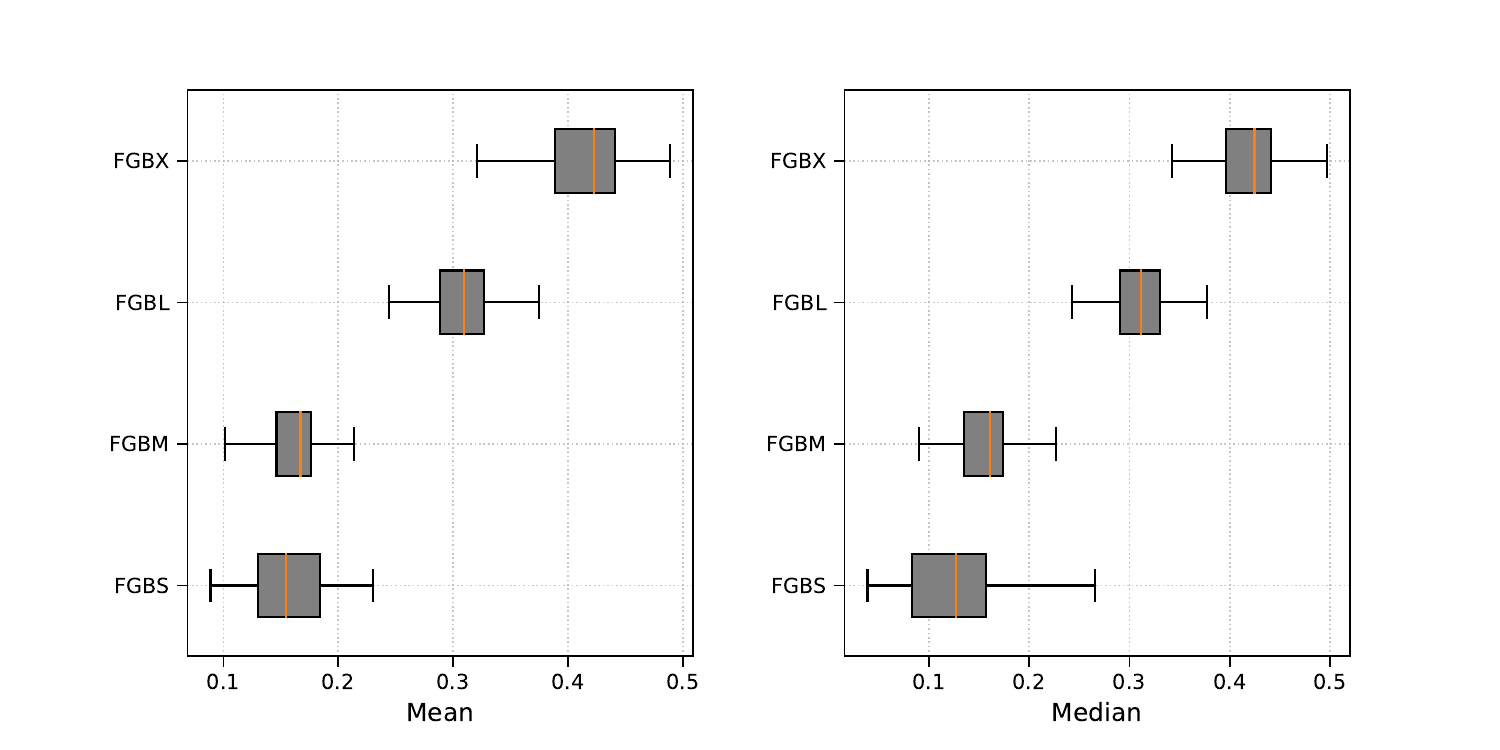}
\caption{Box plot of mean and median correlations between volume and volatility in 10-minute windows.}
\label{fig: volume volatility positive correlation box plot}
\end{figure}

In Figure \ref{fig: volume volatility positive correlation box plot}, a box plot delineates the positive mean and median correlations across multiple days of the data, also showcasing a range of these values for each asset, thus bolstering the evidence of a positive volume-volatility relationship.

These empirical findings support the existence of a positive correlation between volume and volatility in financial markets. The higher trading activity during periods of increased volatility reflects market participants' response to changing market conditions and their desire to adjust positions or capitalize on price movements.

\subsection{Long Range Dependence}

Long-range dependence, also known as long memory, is a stylized fact of financial time series data. It refers to the observation that the data exhibits persistence over extended periods, meaning that the current value of the data is influenced by its past values, even those far in the past.

Mathematically, let $\rho^{\mathrm{abs}}_k$ be the autocorrelation of absolute returns at lag $k$ ($\rho^{\mathrm{abs}}_{k, \Delta t} = \mathrm{Corr}(|R_{t, \Delta t}|, |R_{t-k\Delta t, \Delta t}|)$). The long-range dependence stylized fact can be expressed as the autocorrelation decaying slowly as the lag increases:

\begin{equation*}
\rho^{\mathrm{abs}}_{k, \Delta t} \approx ck^{-\alpha}
\end{equation*}

where $c$ is a constant and $\alpha$ is a parameter determining the rate of decay of the autocorrelation. A value of $\alpha < 1$ indicates long-range dependence, with $\alpha$ closer to zero indicating stronger long-range dependence.

\begin{figure}[H]
	\centering
	\includegraphics[width=\textwidth]{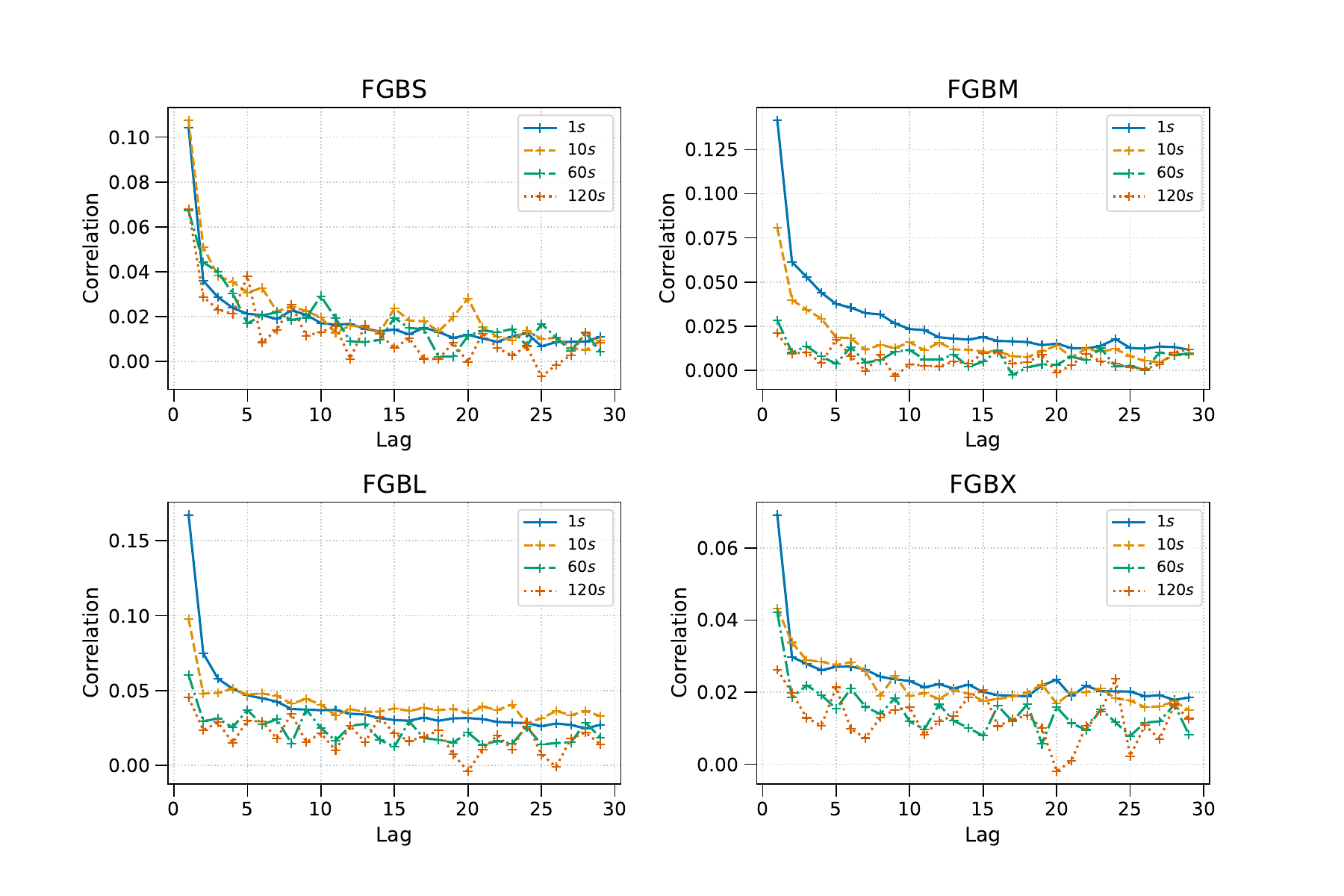}
	\caption{Autocorrelation function of absolute returns for different sampling frequencies.}
	\label{fig:Autocorrelation of absolute returns - mid}
\end{figure}

Figure \ref{fig:Autocorrelation of absolute returns - mid} illustrates the autocorrelation of absolute returns, revealing an apparent power-law decay of the correlation. However, as the sampling frequency increases, the distinct power-law decay shape diminishes, suggesting that this stylized fact holds true primarily for relatively low sampling frequencies ($\leq 10 s$).

\begin{figure}[H]
	\centering
	\includegraphics[width=\textwidth]{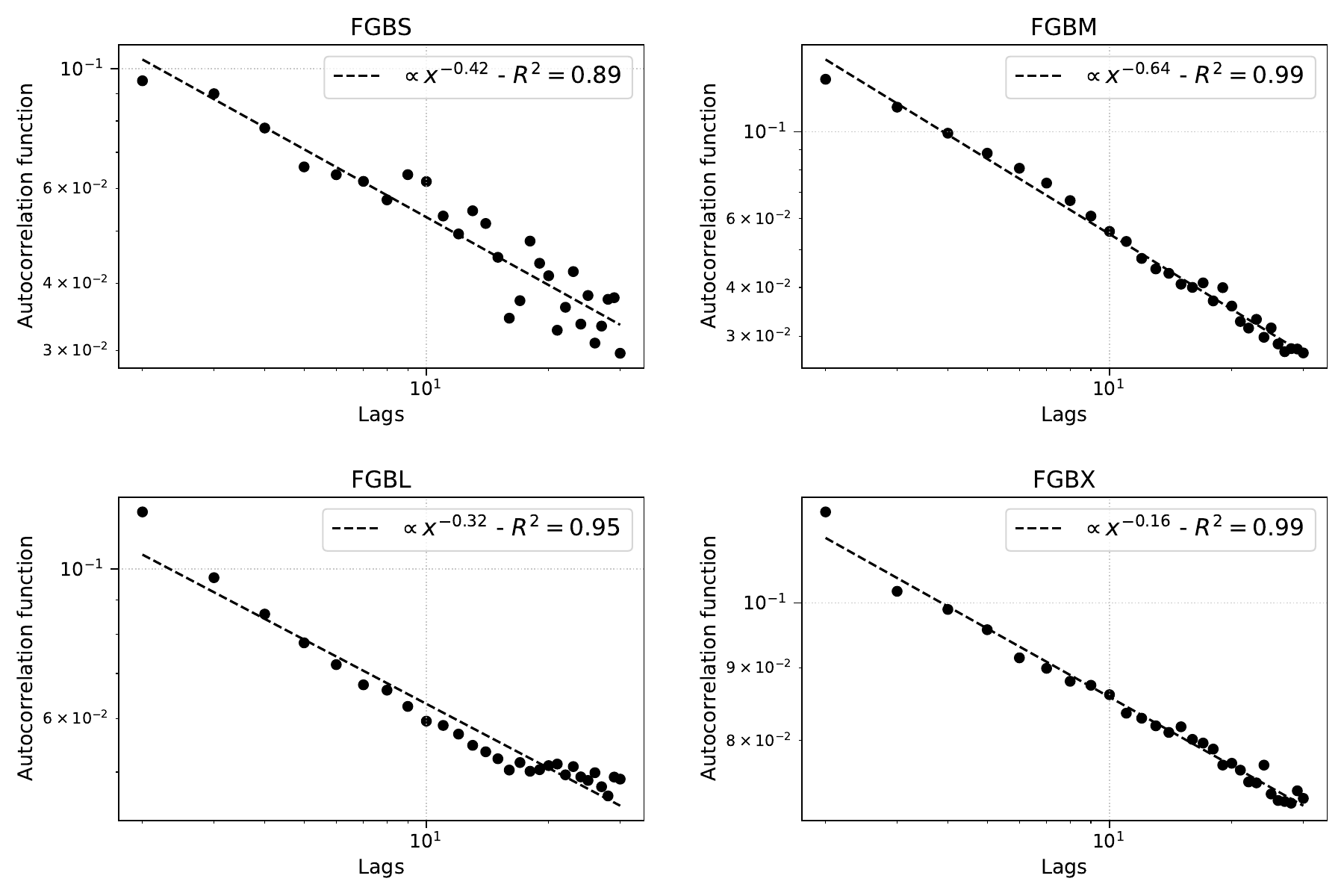}
	\caption{Autocorrelation function and power-law fit (logarithmic scale) for $\Delta t = 1s$.}
	\label{fig:Autocorrelation and fitted line (log-log scale) - sampling frequency = 1 seconds}
\end{figure}

To further investigate the goodness of fit of a power-law distribution, the evolution of $\rho^{\mathrm{abs}}_{k, \Delta t}$ with respect to $k$ is shown in Figure \ref{fig:Autocorrelation and fitted line (log-log scale) - sampling frequency = 1 seconds} for the four instruments. The figure demonstrates that the power-law fit is very good (quite high $R^2$ coefficients), which indicates that the empirical observation is valid for these instruments as well. Moreover, the values of slopes (which is the value of $\alpha$) reveal that long-range dependence is quite strong for all instruments.\\

\begin{figure}[htbp]
	\centering
  	\includegraphics[width=0.8\textwidth]{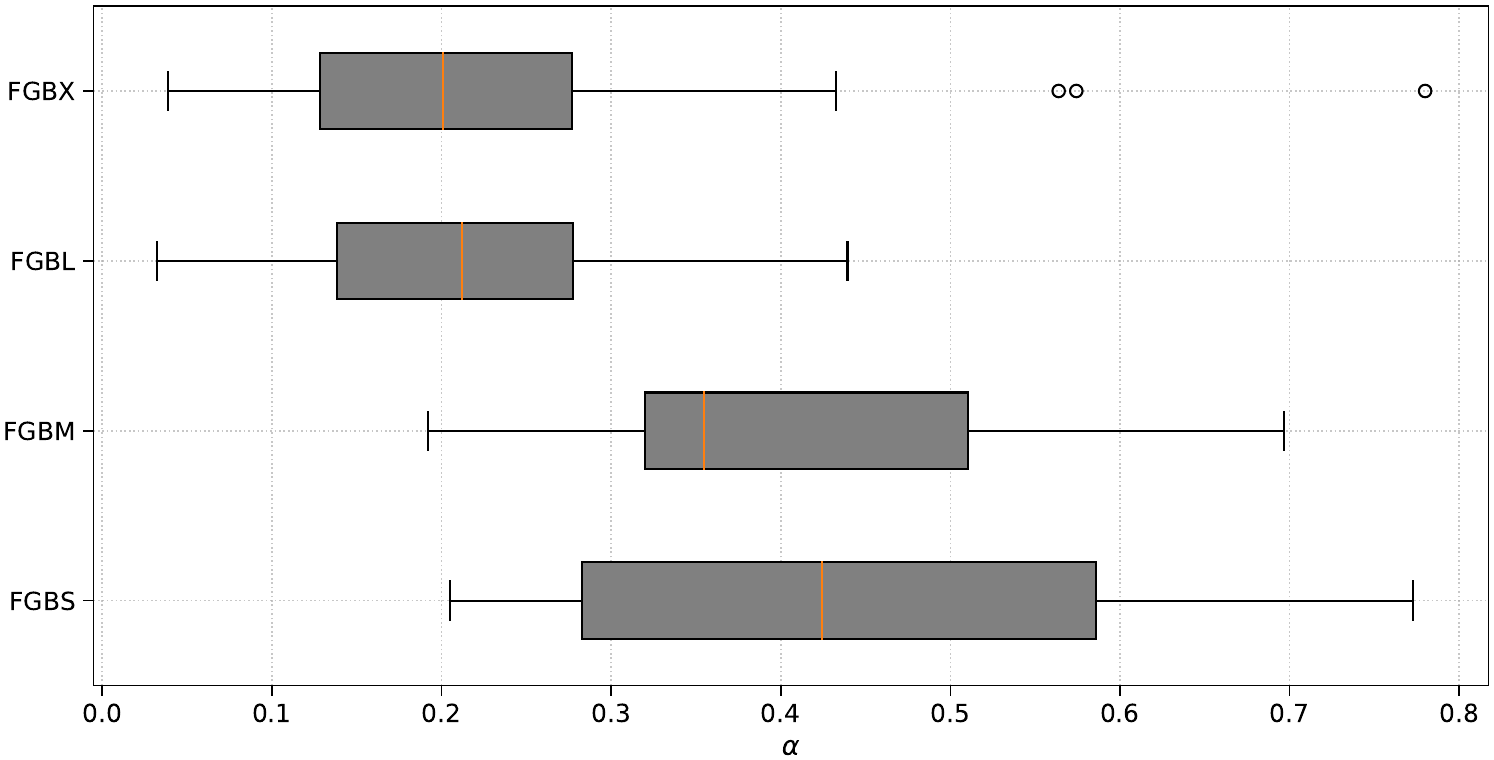}
	\caption{Box plot of fitted $\alpha$ parameter for $\Delta t = 1s$.}
	\label{fig:Long range dependence over days}
\end{figure}

Furthermore, to evaluate the consistency of the fitted parameters, we investigate their value over different days of data. Figure \ref{fig:Long range dependence over days} shows the range of values for each instrument. A similarity can be observed among the instruments in the fitted parameters, and the range of values of $\alpha$, which is all the time less than 1, indicates a strong long-range dependence.

\section{Stylized Facts about Volumes and Order Flow}

\subsection{Order Book Volumes}

The stylized fact regarding order book volumes is based on the observation that the volumes at the best bid ($V_b$) and the best ask ($V_a$) tend to follow a Gamma distribution. Mathematically, it can be expressed as:
$$p(V_a) \propto \exp \left(-\frac{V_a}{V}\right) \left(\frac{V_a}{V}\right)^{\gamma - 1}$$
This equation suggests that the probability distribution of volumes at the best ask conforms to a Gamma distribution, with the shape parameter being $\gamma$ and the scale parameter denoted as $V$.

\begin{figure}[htbp]
	\centering
	\includegraphics[width=\textwidth]{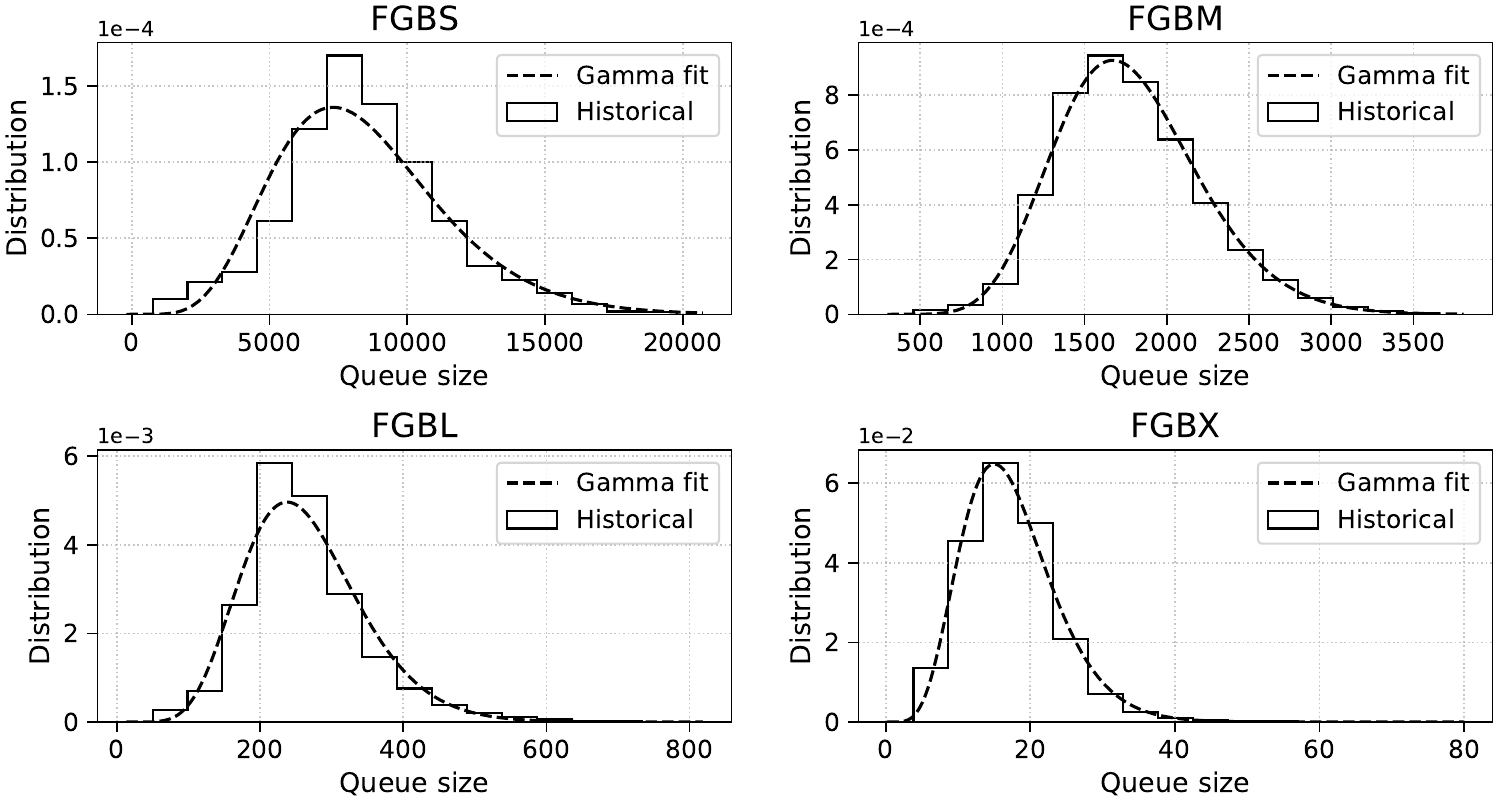}
	\caption{Distribution of volumes available at the best ask along with the fitted Gamma distribution.}
	\label{fig:gamma fit of queue sizes}
\end{figure}

Figure \ref{fig:gamma fit of queue sizes} presents the distribution of volumes at the best ask and the corresponding Gamma distribution fit.\footnote{Similar results were found for the best bid level.} The figure demonstrates that the Gamma distribution provides a good fit for all instruments, with gamma values greater than 1. This finding is consistent with observations in other markets, such as equities \citep{abergel2016limit, vyetrenko2020get}. However, there are differences in the parameters of the fit. For instance, \citet{potters2005financial} find that $\gamma$ is always less than 1 in some equity markets (e.g. France-Telecom and Total). This difference is possibly due to the presence of larger liquidity in the markets subject to our study. Furthermore, this stylized fact observation extends to deeper levels as well.

\begin{figure}[htbp]
	\centering
	\includegraphics[width=0.7\textwidth]{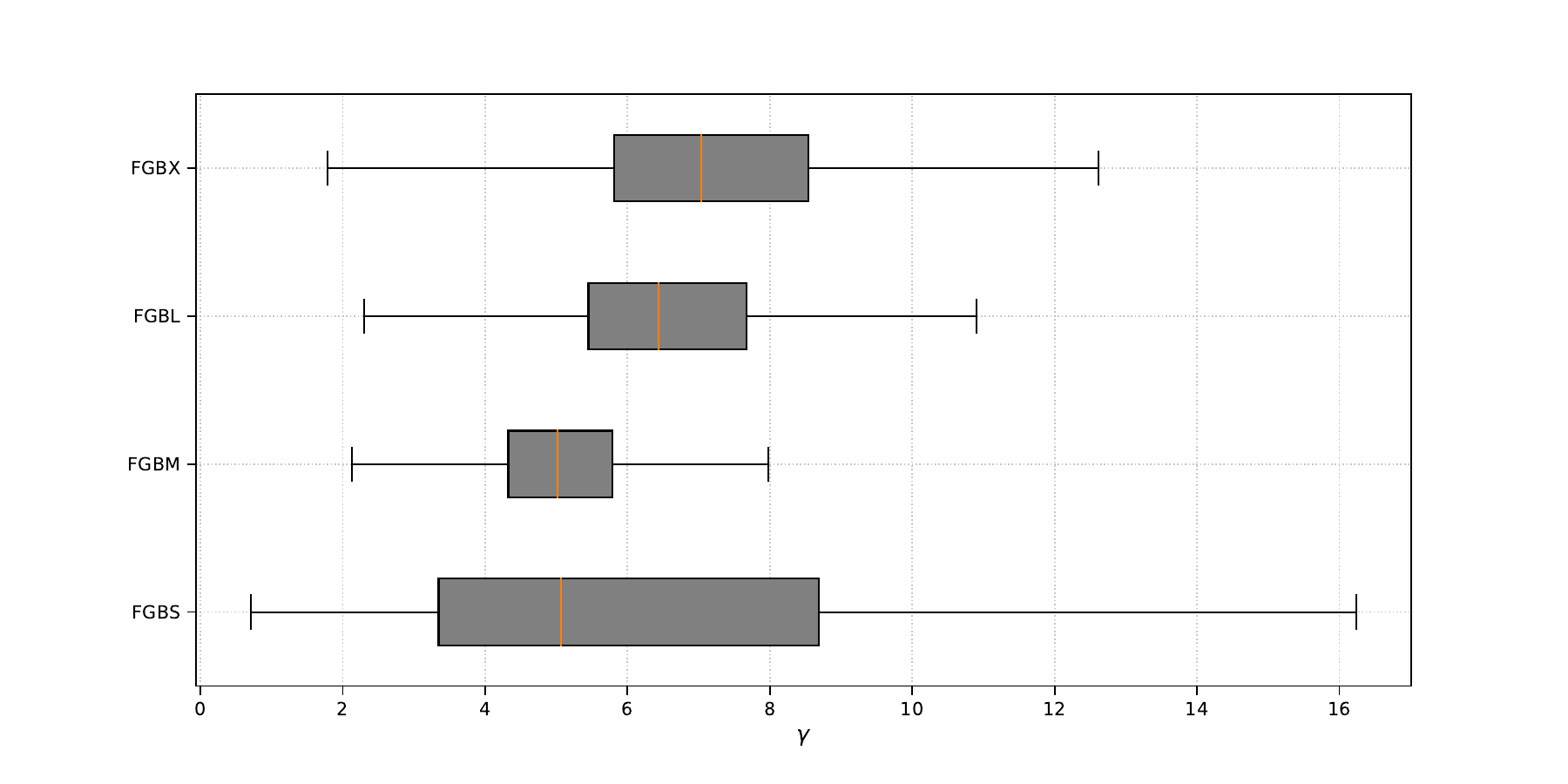}
	\caption{Box plot of the fitted Gamma parameter ($\gamma$).}
	\label{fig:box plot of fitted gamma parameter volumes}
\end{figure}

Figure \ref{fig:box plot of fitted gamma parameter volumes} presents a box plot of the fitted Gamma parameter ($\gamma$), providing a range of gamma values across different instruments. The results demonstrate that gamma values are consistently greater than 1 for all instruments, further supporting the observation that volumes at the best bid and ask to follow a Gamma distribution.

\subsection{Order Sizes Profile}

The order size distribution stylized fact refers to the observation that the sizes of orders placed in limit order books follow a power-law distribution \citep{bouchaud2018trades}. This means that the probability of an order of a given size being placed is proportional to the size of the order raised to a negative exponent, which means that market order sizes tend to be distributed according to the probability distribution:
$$
p(x) \propto x^{-\alpha}
$$

\begin{figure}[H]
	\centering
	\includegraphics[width=\textwidth]{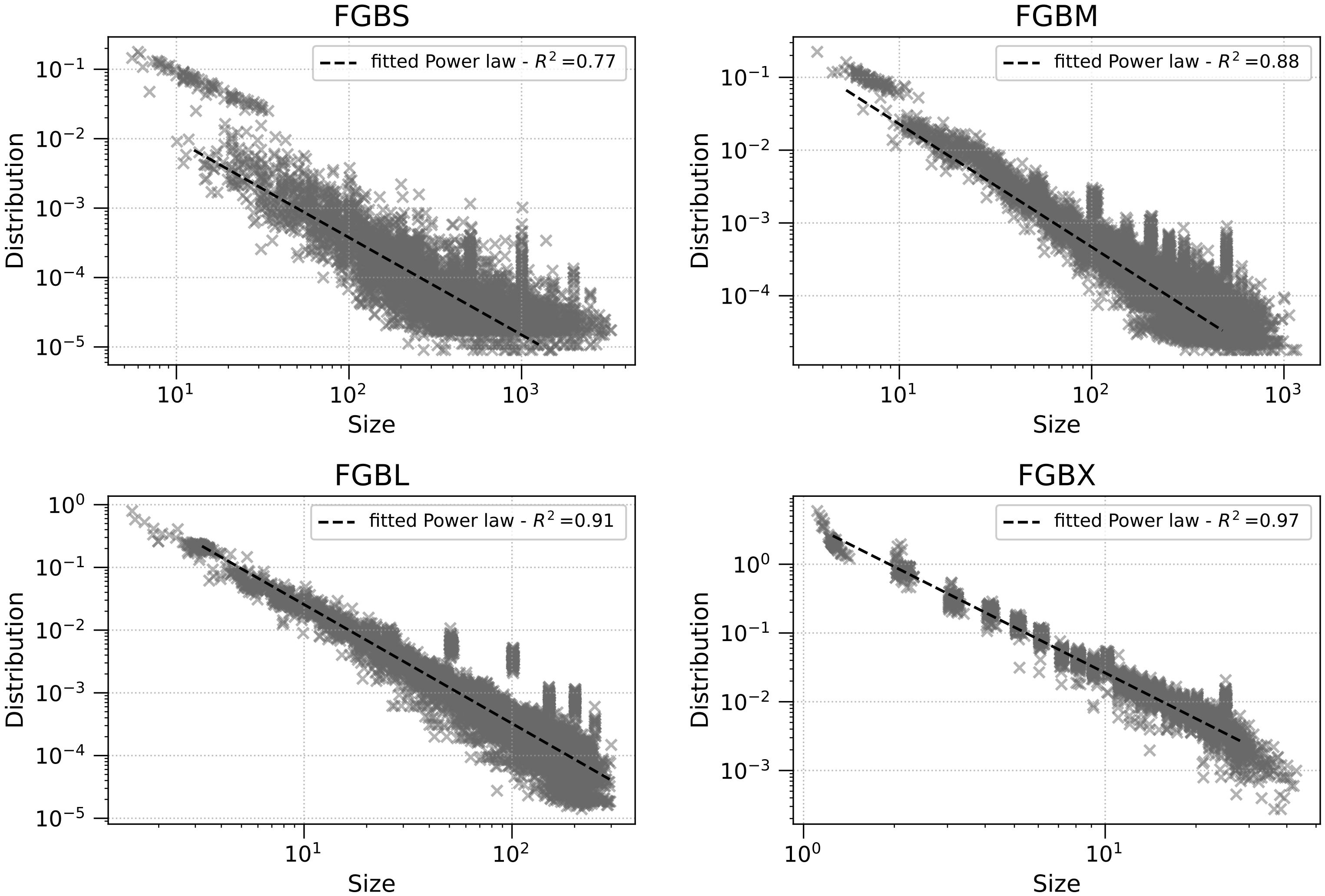}
	\caption{Power-law of order sizes.}
	\label{fig:power-law of order sizes - per instrument}
\end{figure}

Additionally, order sizes often cluster around round numbers of shares, indicating that multiples of 10, 50, 100, etc., are more prevalent than neighboring sizes. Figure \ref{fig:power-law of order sizes - per instrument} depicts the distribution of order sizes (in logarithmic scale) for each instrument, alongside a power-law fit, which displays highly satisfactory fit results (as evidenced by high $R^2$ values). This figure also highlights how these round numbers are particularly prominent and occur more frequently than adjacent size values.

\begin{figure}[htb]
	\centering
	\includegraphics[width=0.8\textwidth]{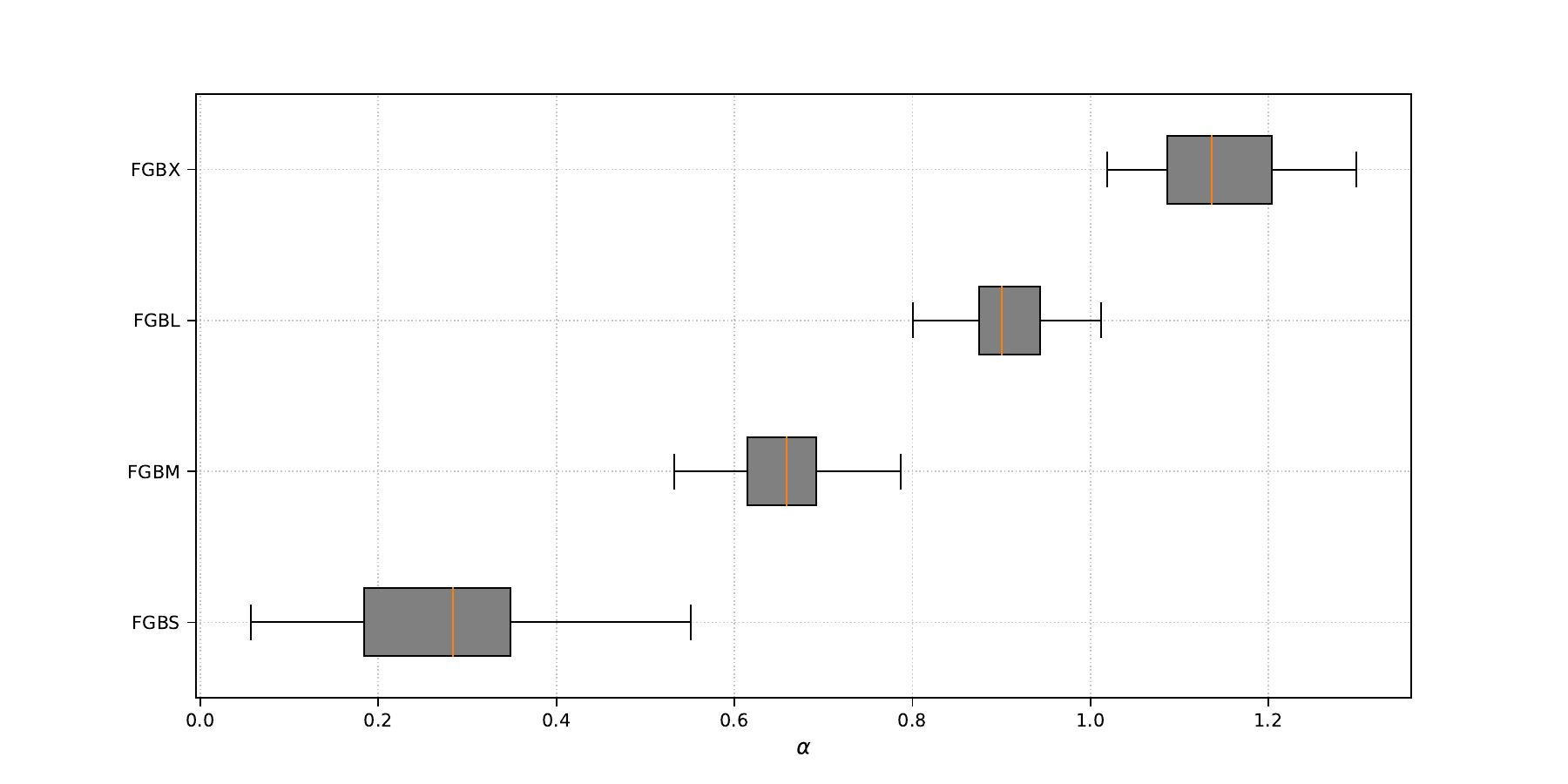}
	\caption{Box plot of fitted power-law parameter.}
	\label{fig:Box plot of fitted powerlaw parameter}
\end{figure}

Figure \ref{fig:Box plot of fitted powerlaw parameter} is the box plot of the fitted $\alpha$ parameter of the power-law distribution. It shows how each instrument has a specific range of values. This difference comes from the unique trading characteristics and liquidity conditions associated with each instrument, which influence the distribution of order sizes. However, the range is quite narrow for each instrument. Furthermore, the figure emphasizes that the exponents of fit increase with the maturity range of the futures. This trend is likely due to the fact that the more long-term a future contract becomes, the more risk it entails, thereby diminishing the willingness of market participants to expose themselves, a behavior that is reflected in the order sizes.

\subsection{Number of Orders in a Fixed Window}

The stylized fact related to order flow posits that the number of orders placed within a fixed time window can be approximated by Gamma or Log-normal distributions \citep{abergel2016limit, vyetrenko2020get}.

\begin{figure}[htbp]
	\centering
	\includegraphics[width=0.95\textwidth]{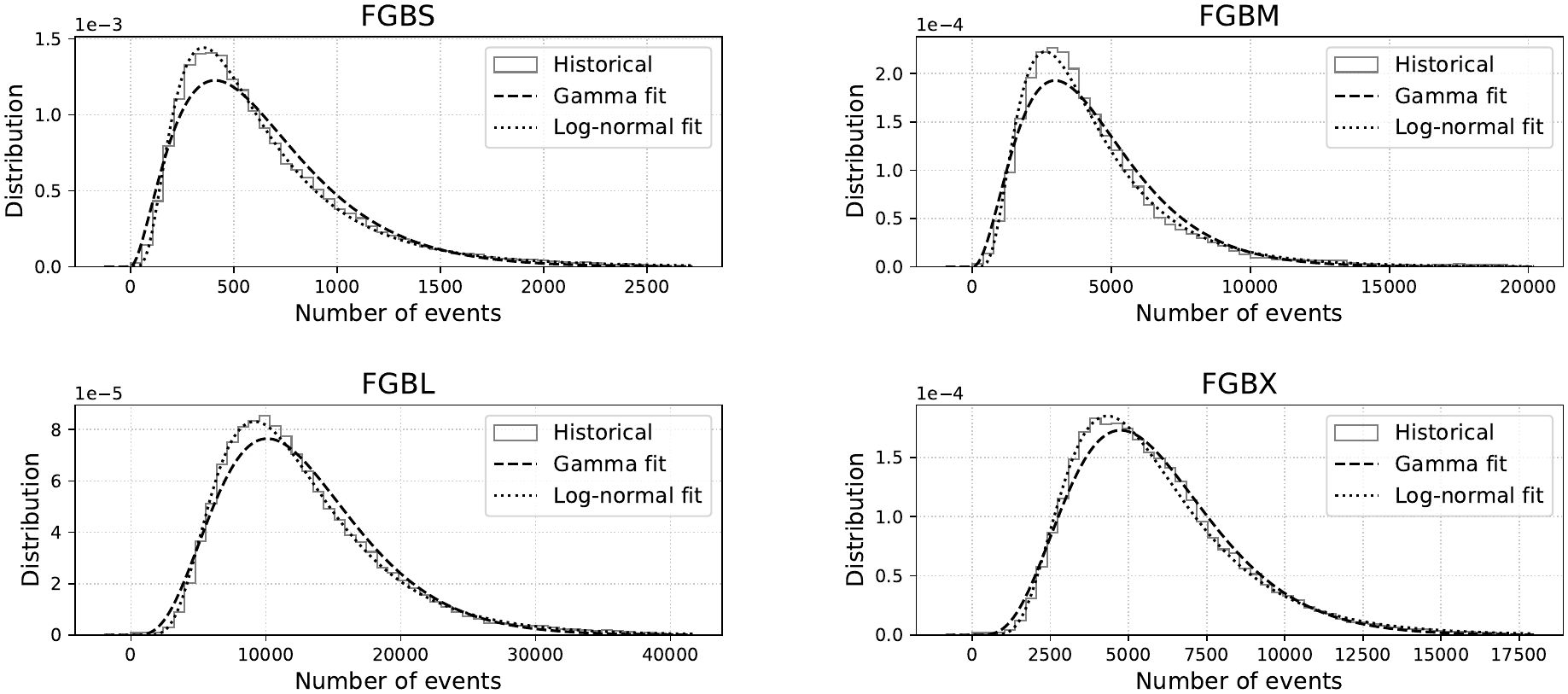}
	\caption{Distribution of the number of events in 5-minute windows and fit of Gamma and Log-normal distributions.}
	\label{fig:orderbook oder count in a given window 300S}
\end{figure}

\begin{figure}[H]
	\centering
	\includegraphics[width=0.95\textwidth]{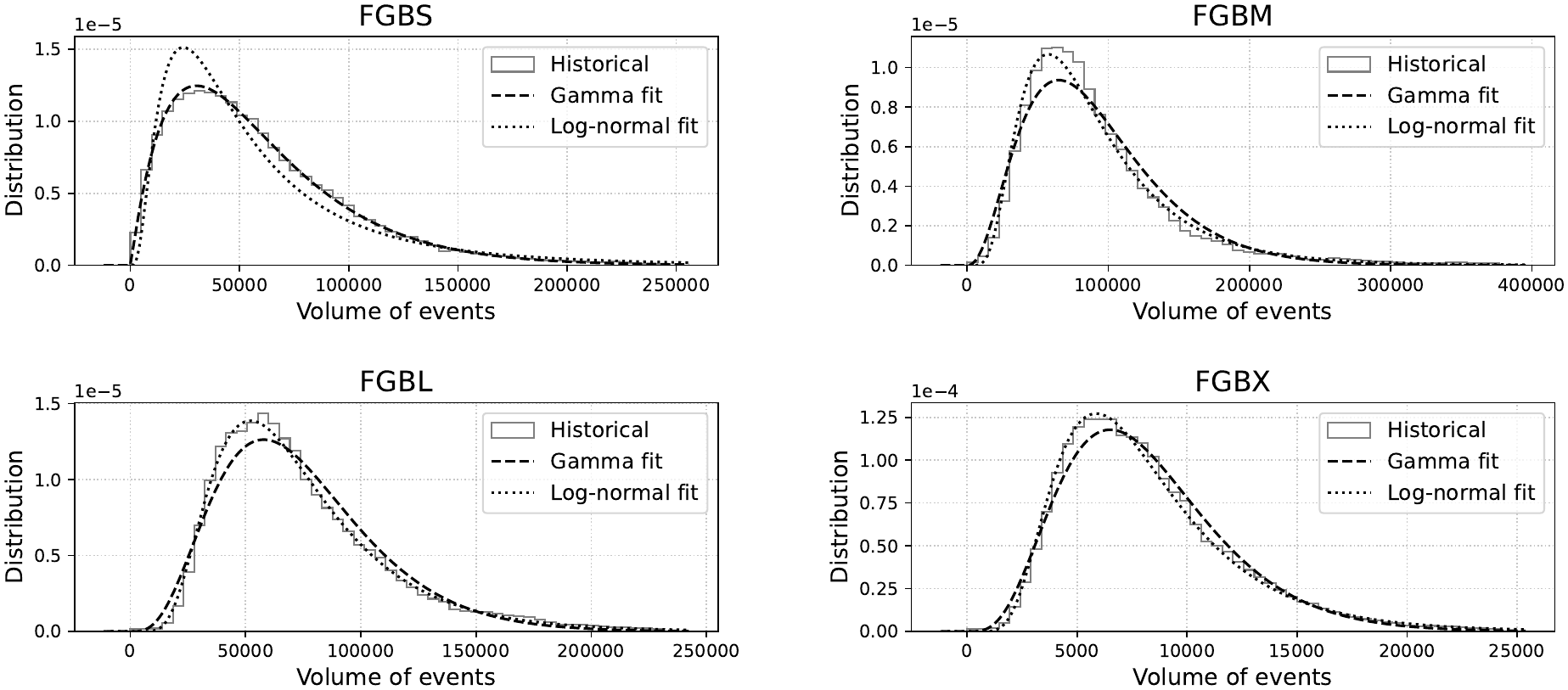}
	\caption{Distribution of aggregated order volumes in 5-minute windows and fit of Gamma and Log-normal distributions.}
	\label{fig:orderbook oder volumes in a given window 300S}
\end{figure}

Figure \ref{fig:orderbook oder count in a given window 300S} presents the empirical distribution of the number of orders in 5-minute windows, juxtaposed with the best fit of each of the two candidate distributions. Here, the Log-normal distributions demonstrate the best fit across all four instruments. This stylized fact is also evident in the distribution of the total volume of orders within a given window, as depicted in Figure \ref{fig:orderbook oder volumes in a given window 300S} that illustrates that at least one of the two distributions aligns remarkably well with the empirical data.

\newpage
\subsection{Interarrival Time of Orders}

The interarrival time stylized fact refers to the observation that the time between subsequent orders in the limit order book can be fit into various probability distributions. Specifically, market order book interarrival times can be fit into Exponential, Log-normal, and Weibull distributions \citep{abergel2016limit}. This means that the time elapsed between two subsequent market orders\footnote{In the contemporary market, there has been a discernible decline in the utilization of market orders, with a corresponding increase in the adoption of aggressive limit orders as a preferred trading strategy. This notable shift can be attributed to a multitude of reasons. Firstly, the implementation of limit orders grants investors the ability to determine the maximum or minimum price at which they are willing to engage in the purchase or sale of a given security. This affords them a greater degree of control over the execution price. Moreover, aggressive limit orders entail significantly smaller opportunity costs when compared to market orders. Consequently, investors can potentially obtain more favorable prices by exercising patience and awaiting market conditions that align with their specified price \citep{Handa1995Limit, Chou2009Strategic}. For the sake of simplicity, we consider all orders leading to transactions as equivalent to market orders.
}
in a limit order book can follow one of these distributions, depending on the market conditions and the asset being traded.

\begin{figure}[htbp]
	\centering
	\includegraphics[width=\textwidth]{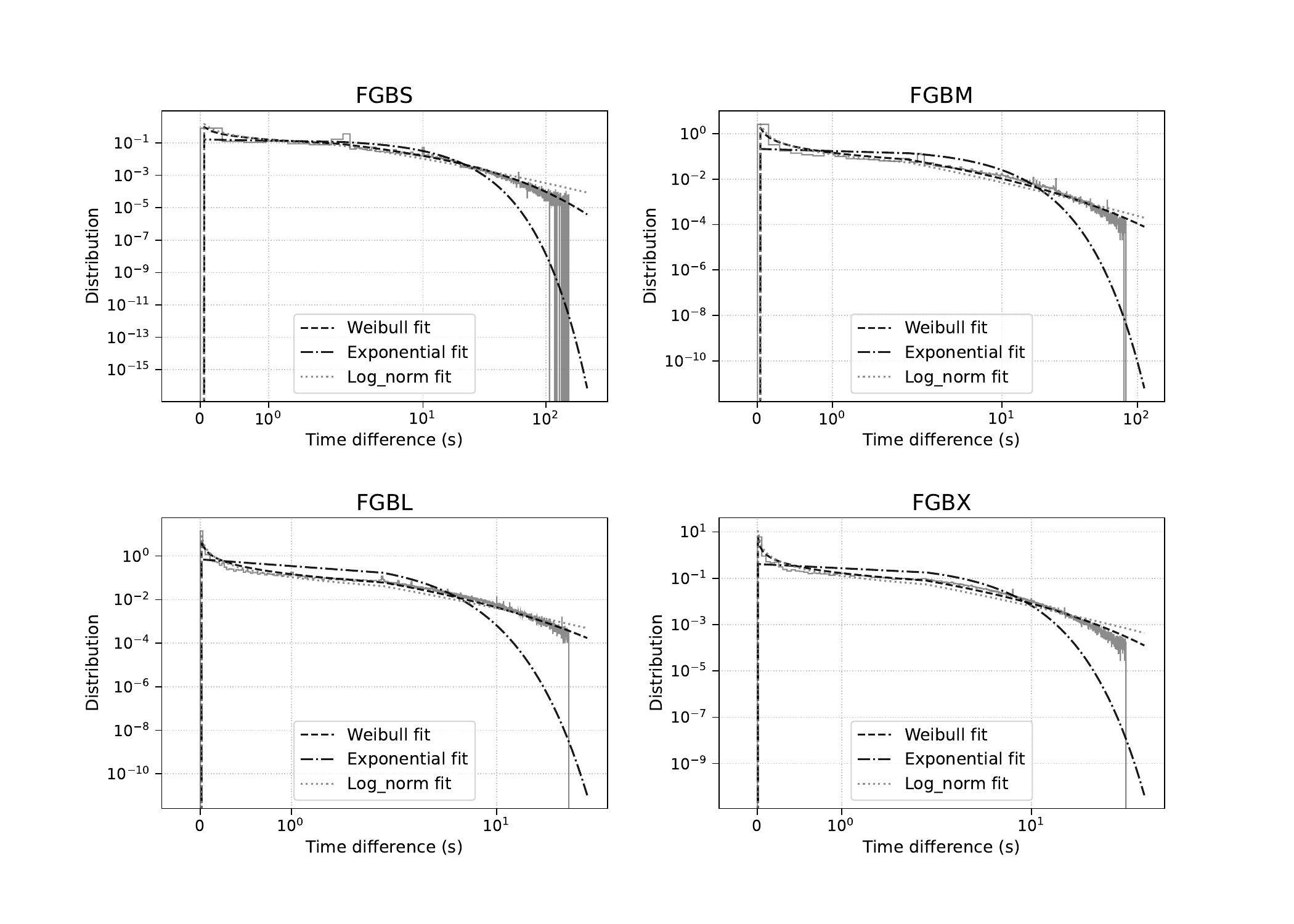}
	\caption{Distribution of interarrival time between orders and Weibull fit.}
	\label{fig:wiebull fit or interarrival of events - physical time}
\end{figure}

This observation has been investigated for different time types, and similarly to \citet{abergel2016limit}, the Weibull distribution is the one to exhibit the best fit for the four instruments, as one can deduce from Figure \ref{fig:wiebull fit or interarrival of events - physical time}. \\

\begin{figure}[htbp]
	\centering
	\includegraphics[width=\textwidth]{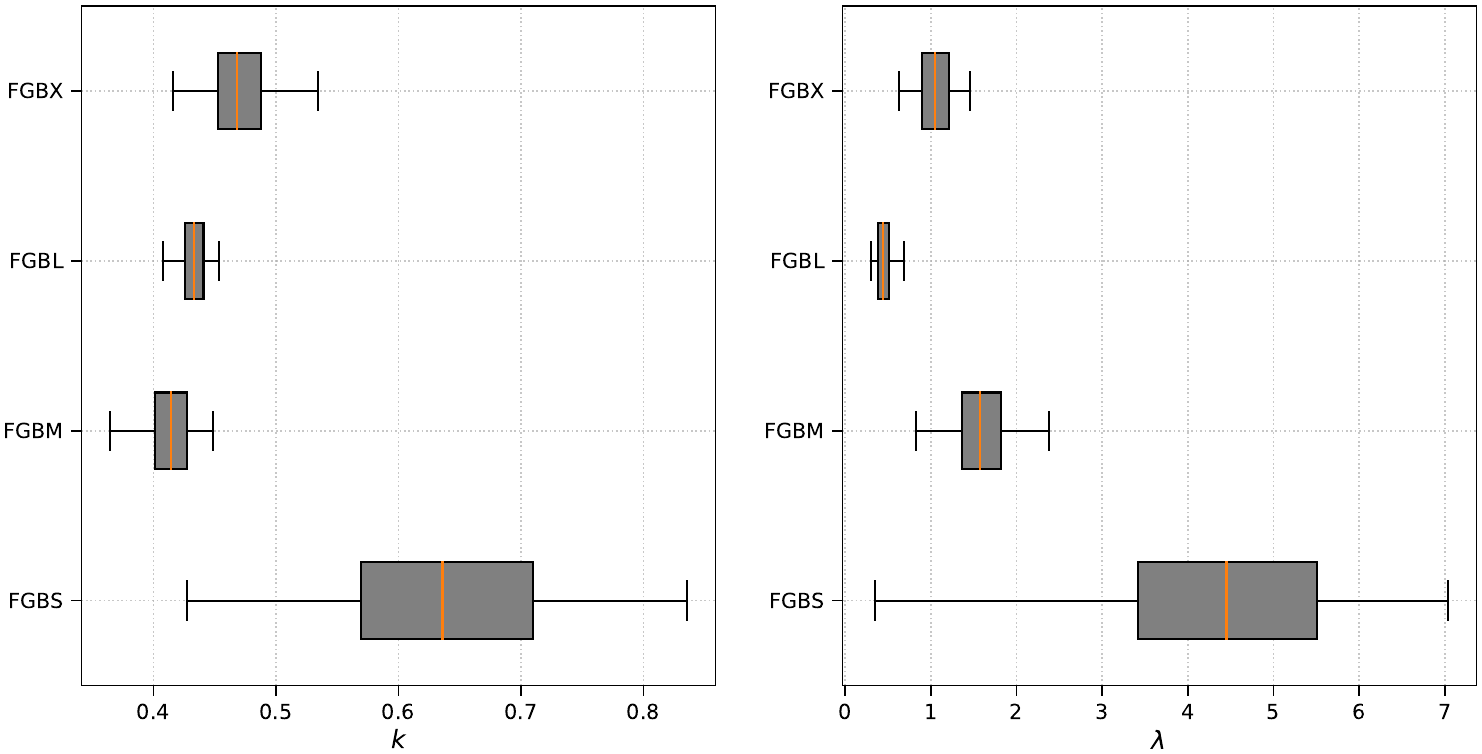}
	\caption{Box plot of fitted Weibull parameters.}
	\label{fig:wiebull fit or interarrival of events - multiple days - physical time}
\end{figure}

In fact, for $k, \lambda \in \mathbb{R}^*_+$, the Weibull distribution density function is defined as:

$$
f_{(k, \lambda)}(x)= \begin{cases}\frac{k}{\lambda}\left(\frac{x}{\lambda}\right)^{k-1} e^{-(x / \lambda)^k}, & x > 0 \\ 0, & x\leq 0\end{cases}
$$

where $k \in \mathbb{R}^*_+$ is the shape parameter and $\lambda \in \mathbb{R}^*_+$ is the scaling parameter.

Figure \ref{fig:wiebull fit or interarrival of events - multiple days - physical time} shows the box plot of fitted parameters $k$ and $\lambda$ over different days of the data. It shows that these parameters lie in a specific interval that depends on the instrument, with a quite narrow interval for FGBL, FGBM, and FGBX, in contrast with FGBS which has a very wide one. This could be due to the unique characteristics of FGBS, such as its shorter maturity period and higher market order rates, which could lead to more rapid changes in its price and hence a wider range of fitted parameters.

\begin{figure}[H]
	\centering
	\includegraphics[width=0.7\textwidth]{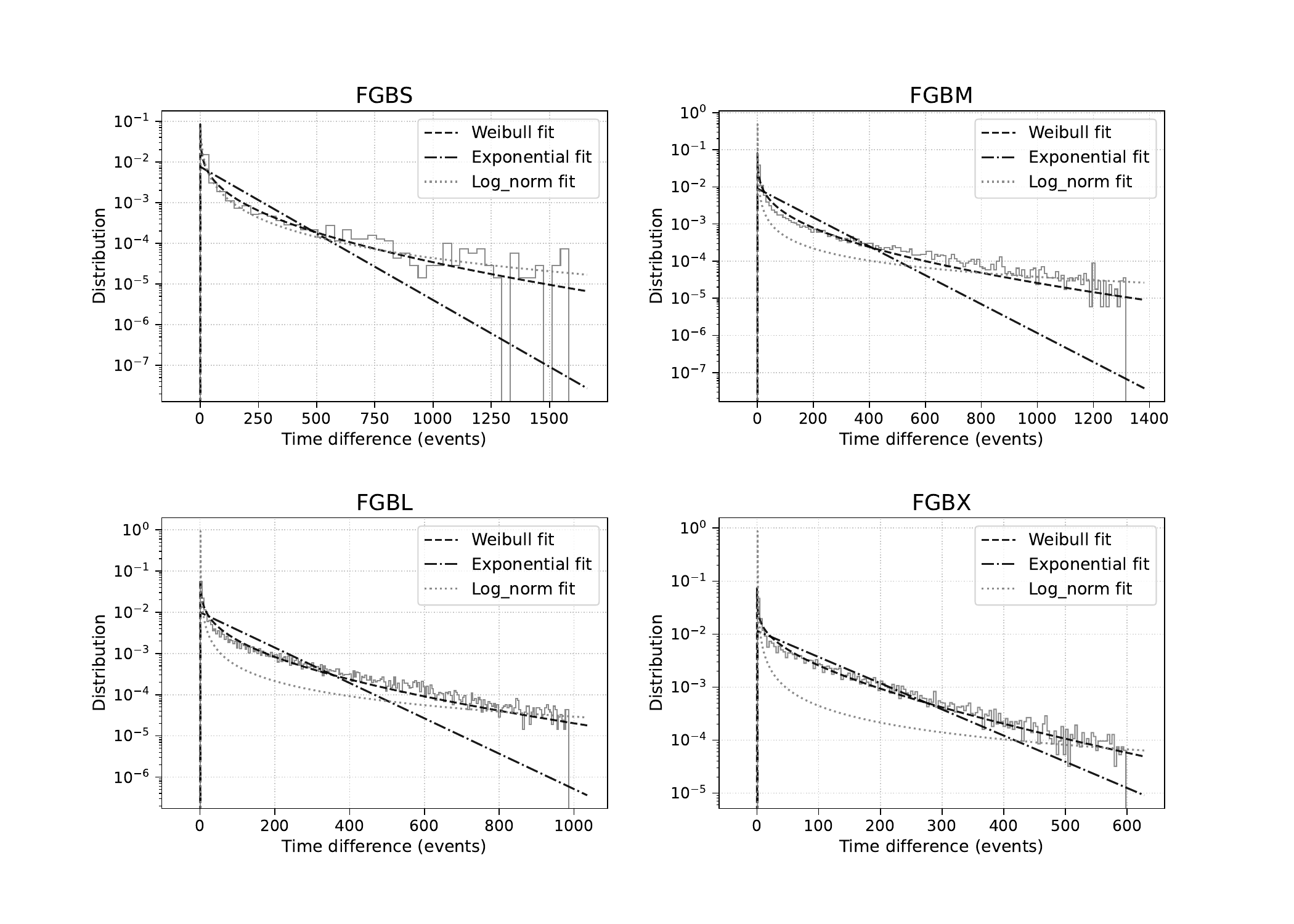}
	\caption{Distribution of interarrival time between market orders and Weibull fit - Event time.}
	\label{fig:wiebull fit or interarrival of events - event time}
\end{figure}

\begin{figure}[H]
	\centering
	\includegraphics[width=0.6\textwidth]{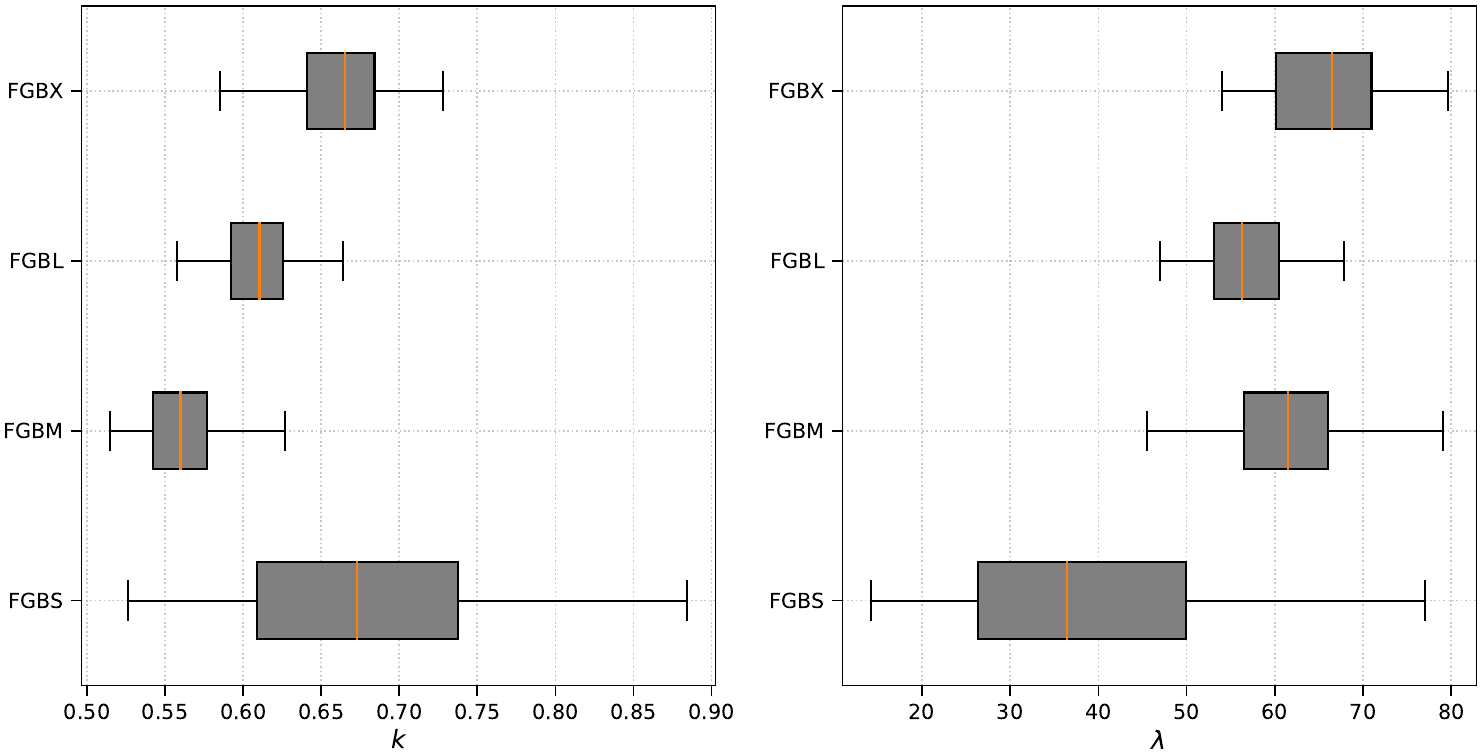}
	\caption{Box plot of fitted Weibull distributions - Event time.}
	\label{fig:wiebull fit or interarrival of events - multiple days - event time}
\end{figure}

Furthermore, we investigated the validity of such a fit for other types of times. Particularly, the event time shows similar behavior, where the Weibull distribution exhibits a good fit for the four instruments (Figure \ref{fig:wiebull fit or interarrival of events - event time}), and the fitted parameters of this distribution show consistency over different days of the data (Figure \ref{fig:wiebull fit or interarrival of events - multiple days - event time}).

In conclusion, the Weibull distribution provides the best fit for the four assets, and the fitted distribution closely follows the historical distribution. Furthermore, the fitted parameters demonstrate homogeneity between data from the same instrument, with FGBL, FGBM, and FGBX showing similar behavior and overlapped range of values. Schatz future, however, exhibits different behavior, with larger parameters indicating that the interarrival time of orders is generally longer for this asset.

\subsection{Excitation between Events}

The limit order book is a transparent platform where Market Participants (MPs) can see the state of the market and the overall behavior of other participants. This causes a mutual exchange of information, which is reflected in the way that MPs respond to past events. For example, the behavior of MPs when a market order arrives is different than when a limit order arrives, even more, when the side of the event changes.

\begin{figure}[htbp!]
	\centering
	\includegraphics[width=0.85\textwidth]{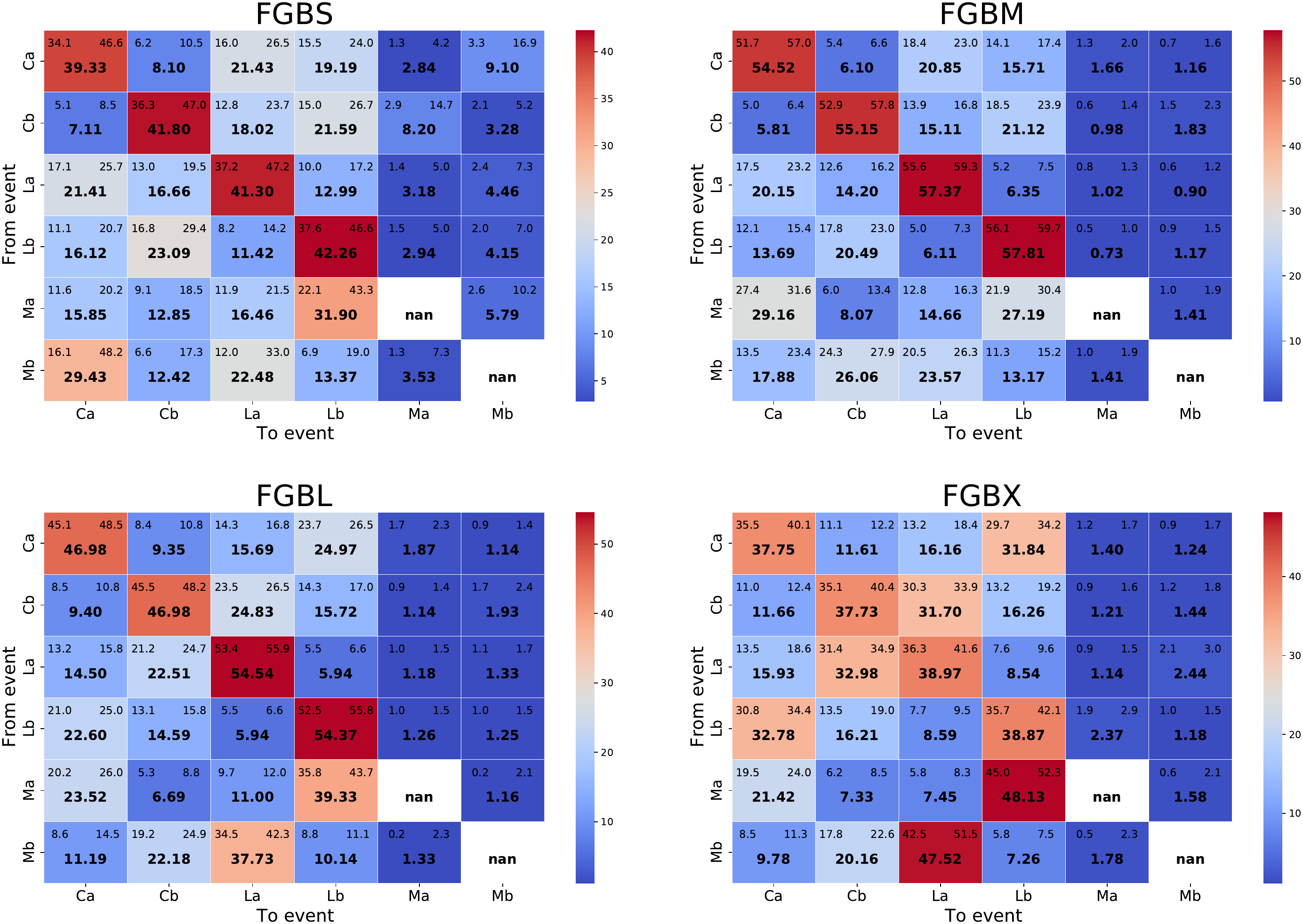}
	\caption{Transition matrix of event types and sides. First letter indicates the type of event (C for \emph{cancel}, L for \emph{limit} and M for \emph{market}, second letter indicates the side of the order, \emph{a} for ask and \emph{b} for bid).}
	\label{fig:Excitation matrix bid ask   probabilities}
\end{figure}

Particularly, the conditional distribution of events on past events can clearly highlight such an effect. Figure \ref{fig:Excitation matrix bid ask   probabilities} shows the matrix of conditional probabilities of an event $E_{t^+}$ (columns) conditional on past ones $E_{t^-}$ (rows). This means, it states the probabilities $ \probP \left( E_{t^+} | E_{t^-} \right) $. This matrix has been computed for each day and the figure states the average values of these probabilities, flanked by the 5th and 95th percentile values. The figure shows a strong similarity of these values among different days of data of each instrument, and also among assets.

\begin{figure}[htbp!]
	\centering
	\includegraphics[width=0.85\textwidth]{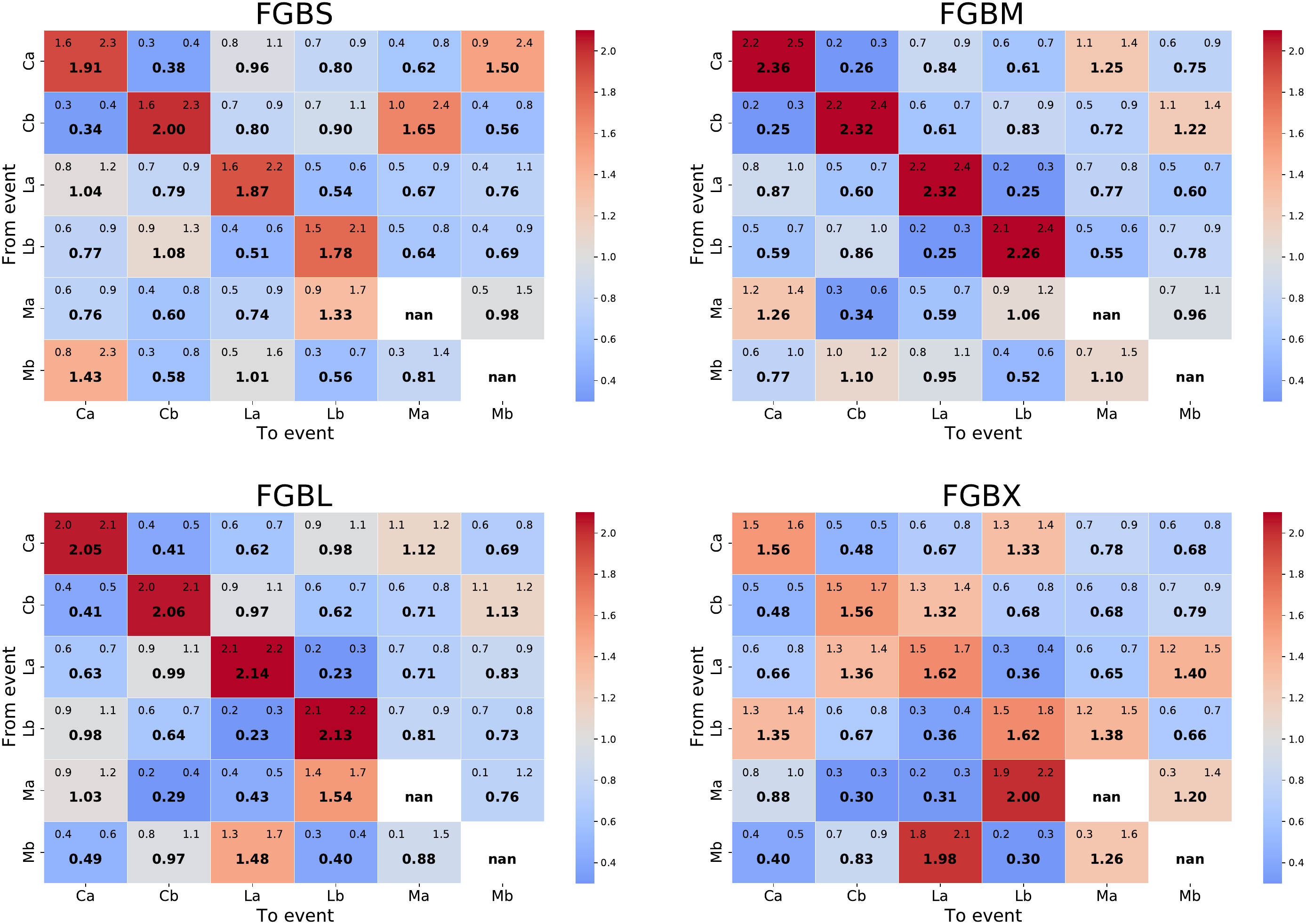}
	\caption{Excitation matrix of event types and sides.}
	\label{fig:Excitation matrix bid ask   excitations}
\end{figure}

Moreover, it is insightful to examine the values of the quantities $ \frac{\probP \left( E_{t^+} | E_{t^-} \right)}{\probP(E_{t^+})} $, which indicate how the event $E_{t^-}$ influences the occurrence of event $E_{t^+}$. When this quantity is significantly greater than 1, it suggests a strong excitation, whereas values significantly less than 1 indicate strong inhibition, meaning that the event $E_{t^-}$ discourages the occurrence of event $E_{t^+}$. \\Figure \ref{fig:Excitation matrix bid ask   excitations} displays these quantities computed for different days, presenting the average value, 5th percentile, and 95th percentile values.

These two figures provide insights into how the three elementary events (limit, cancel and market) and side of order  (ask or bid) influence each other on best limits. Notably, we can observe:

\begin{itemize}
\item High values on the diagonal of the matrix, compared to other values in the same column, unveil the auto-excitation nature of the event. Occasionally, excitation values exceed 2, implying the event occurs at least twice as frequently following a similar event. However, auto-excitation values of market orders have been intentionally omitted due to the nuances in how market orders are recorded in our data, which exclusively contain \emph{splitted orders}. These are market orders that facilitate the asset exchange between two agents and are divided into several smaller orders. This division occurs when the best limit, instead of representing a single, unified order, is an amalgamation of quantities from multiple agents at the top of the order book. Consequently, when a single market order arrives, it is often fragmented into several smaller orders. Each of these smaller orders is matched with a portion of the available quantities at the best limit. This fragmentation process can potentially amplify the perceived market activity artificially. 
As our dataset only provides L2 level data, with no details about the market participants initiating orders, we are unable to aggregate these child orders. Thus, we chose to omit the auto-excitation between market orders on the same sides. Nonetheless, it is our supposition that these values should also be high, considering that realized transactions furnish invaluable insights into market direction and can substantially influence subsequent transactions.

\item Mutual inhibition between limit orders of one side and market orders of the same side. This could be because limit orders and market orders serve different trading strategies. Limit orders are used when traders are not in a hurry to execute a trade and are waiting for a specific price, while market orders are used when quick execution is prioritized over the trade price. This insight underscores the contrasting motivations between those placing limit orders and those placing market orders, reflecting underlying differences in trading strategies and risk tolerances.

\item Cancel orders on one side excite market orders of the same side. This is likely because the rush to cancel an order may signal others to rush for liquidity before the price changes. Additionally, cancel orders on one side excite limit orders on the opposite side more than they do on the same side. For example, the values of excitation of a cancel order on the bid side are greater for a limit order on the ask side than on the bid side. This holds true also for the relationship of market orders towards limit orders. This can be indicative of a more tactical market response. MPs might perceive a cancellation on one side as a lack of confidence in that direction of the market, prompting them to place limit orders on the opposite side to capitalize on the anticipated price movement. This pattern suggests that MPs are continually reading and reacting to the signals provided by others' actions, underscoring the interconnected and strategic nature of market behavior. The same logic can extend to the relationship between market orders and limit orders, where immediate market orders may indicate a strong belief in the current market direction, stimulating some market participants to position themselves on the opposite side, expecting the market to revert.

\end{itemize}

This observed behavior of market participants aligns closely with the findings in \citet{jaisson2015limit} study on the endogeneity of financial markets. The paper notes that empirical measures of Hawkes processes kernels are often close to one, indicating a high degree of endogeneity in financial markets. This suggests that a significant proportion of market orders are endogenously triggered by past orders rather than being exogenous.

\subsection{Intraday Seasonality}

In financial markets, it is well-documented that trading activity is not uniformly distributed throughout the day. This key stylized fact has been observed across various markets, including equities. The intraday pattern of trading activity typically follows a ``U-shape'' \citep{biais1995empirical} characterized by high activity levels at the beginning and end of the trading day, with a quieter period around midday \citep{bouchaud2018trades, vyetrenko2020get}. This pattern is often attributed to the strategic behavior of traders who prefer to trade at specific times of the day, such as the opening and closing periods when liquidity is typically higher and information asymmetry is lower.\footnote{The increase in trading activity, particularly at the opening and closing of the market, might both respond to and also impact prevailing conditions. When traders, especially those managing large positions, become active in these periods, their actions could naturally enhance liquidity, attracting more market participants. Consequently, traders activity and market conditions might exist in a mutually reinforcing relationship.} Furthermore, \citet{abergel2016limit} have successfully fitted a quadratic function to this pattern, further confirming its ``U-shape'' nature.

\begin{figure}[H]
	\centering
	\begin{subfigure}[b]{0.68\textwidth}
		\includegraphics[width=\textwidth]{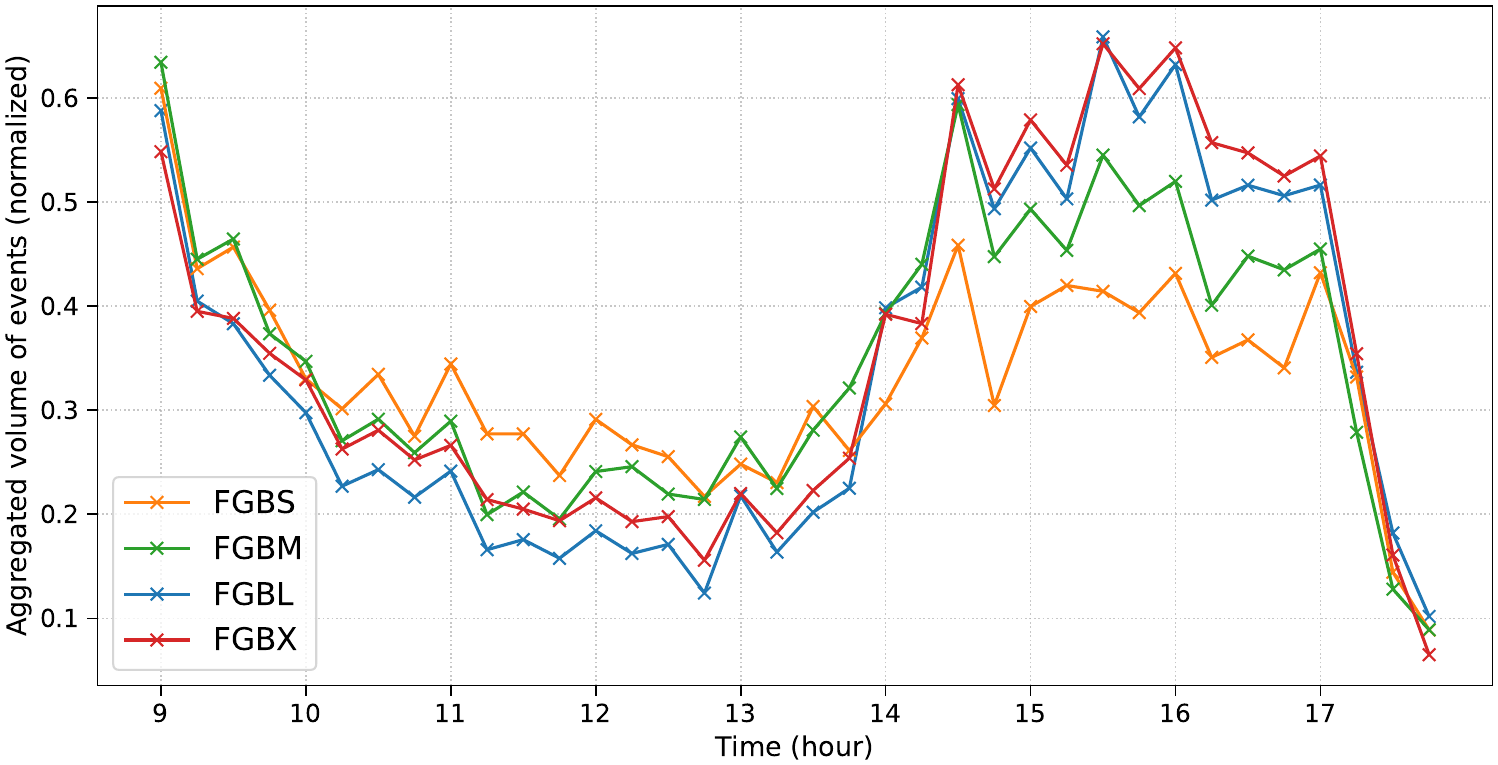}
		\caption{Normalized total volume of orders in windows of size 15 minutes.}
		\label{fig:Volume of events (normalized)}
	\end{subfigure}
	\hfill 
	\begin{subfigure}[b]{0.68\textwidth}
		\includegraphics[width=\textwidth]{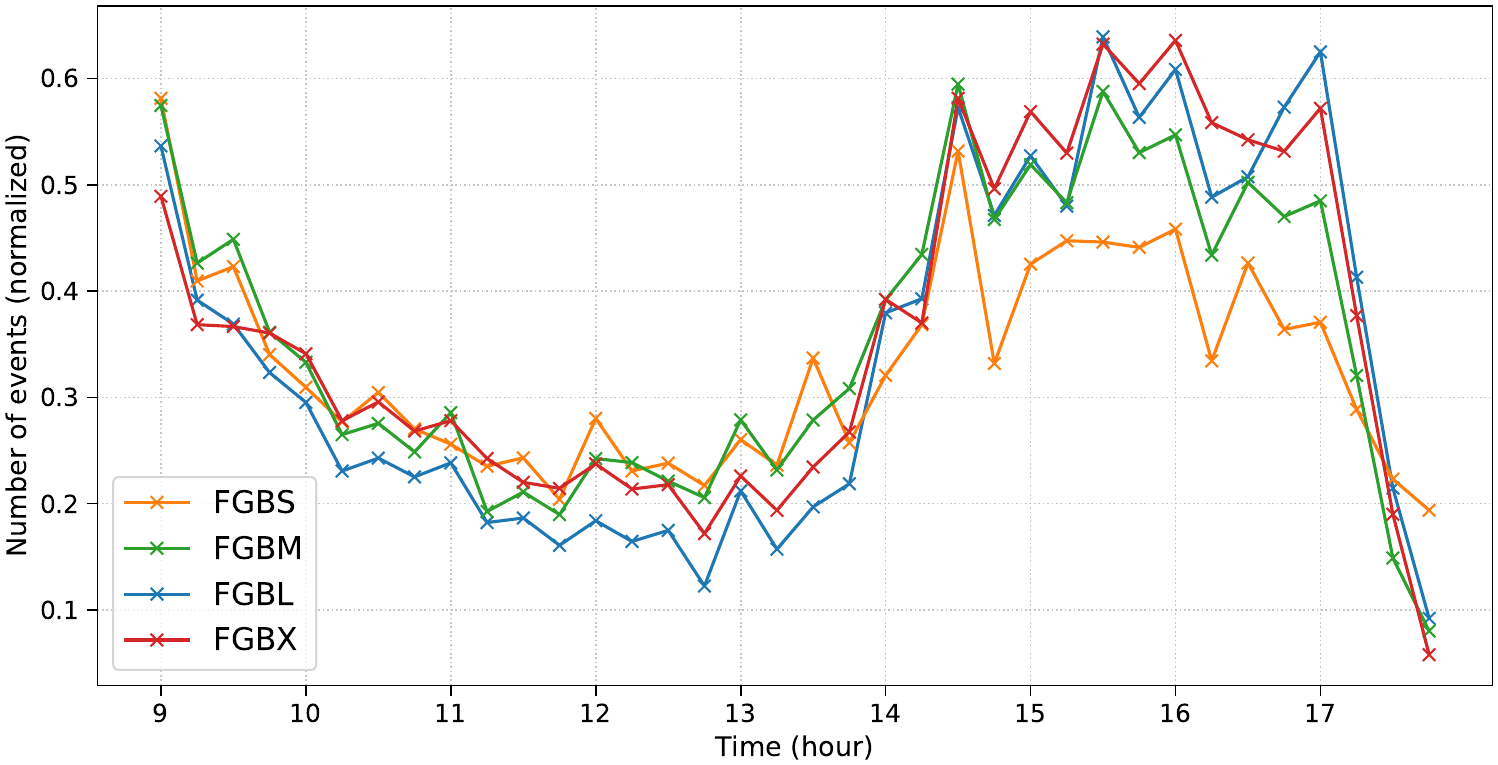}
		\caption{Normalized number of orders in windows of size 15 minutes.}
		\label{fig:Number of events (normalized)}
	\end{subfigure}
	\caption{Normalized volume (a) and number of orders (b) in windows of size 15 minutes.}
\end{figure}

Figures \ref{fig:Volume of events (normalized)} and \ref{fig:Number of events (normalized)} show respectively the normalized total volume of orders and the number of orders in windows of size 15 minutes along active trading hours. These figures exhibit a common pattern observed for the four instruments. The observed trading pattern, characterized by a sharp spike at market open, a decrease until midday, a steady increase until mid-afternoon, and a final spike followed by a rapid decrease at market close, is a common phenomenon in financial markets. This pattern can be interpreted as:

\begin{itemize}
	\item The spike in trading activity observed at 9:00 corresponds to the opening of the market. This is a common phenomenon across many financial markets, as traders respond to overnight news and events, leading to a surge in trading orders that get executed as soon as the market opens.
\item The subsequent decrease in trading activity until 13:00 could be attributed to the diminution of the initial rush of trading activity and the lunch break period, during which many traders pause their trading activities.

	\item The minor spike at 13:00 and the increase in trading activity until 14:30 could be due to traders returning to the market after the lunch break and reacting to any news or events that occurred during the break.
	\item The relatively constant and high trading activity between 14:30 and 17:00 is likely due to this period being one of the most active trading periods of the day, particularly because it overlaps with the opening hours of the US markets. The simultaneous opening of multiple major markets can lead to increased trading activity.
\item The minor spike at 17:00 and the sharp decrease in trading activity afterward could be elucidated by the procedure of determining the daily settlement prices. On the Eurex exchange, the daily settlement prices for the current maturity month are extracted from the volume-weighted average of the prices of all transactions during the minute before 17:15 CET \citep{Eurex_benchmark}, given that more than five trades are transacted within this period. This phenomenon could foster increased trading activity, as traders might opt to execute trades at the settlement price. This could be driven by (i) it being a notably liquid moment—precipitating a sort of self-fulfilling liquidity due to concurrent market participation by numerous traders—and (ii) an endeavor to prevent discrepancies between accounting and mark-to-market (MTM) valuations by transacting at the universally acknowledged settlement price.
\end{itemize}

\section{Stylized Facts about Price Microstructure}

\subsection{Signature Plot}

The signature plot of a price is defined as the variance of price increments for a given lag value $h$, normalized by $h$; mathematically, it is defined as the function $\sigma^2_{h} = \frac{\mathbb{V}\left(m_{t+h} - m_t\right)}{h}$. Tracking the variation of this metric in relation to $h$ offers valuable insights into price dynamics over different time scales.

\begin{figure}[htbp]
	\centering
	\includegraphics[width=0.9\textwidth]{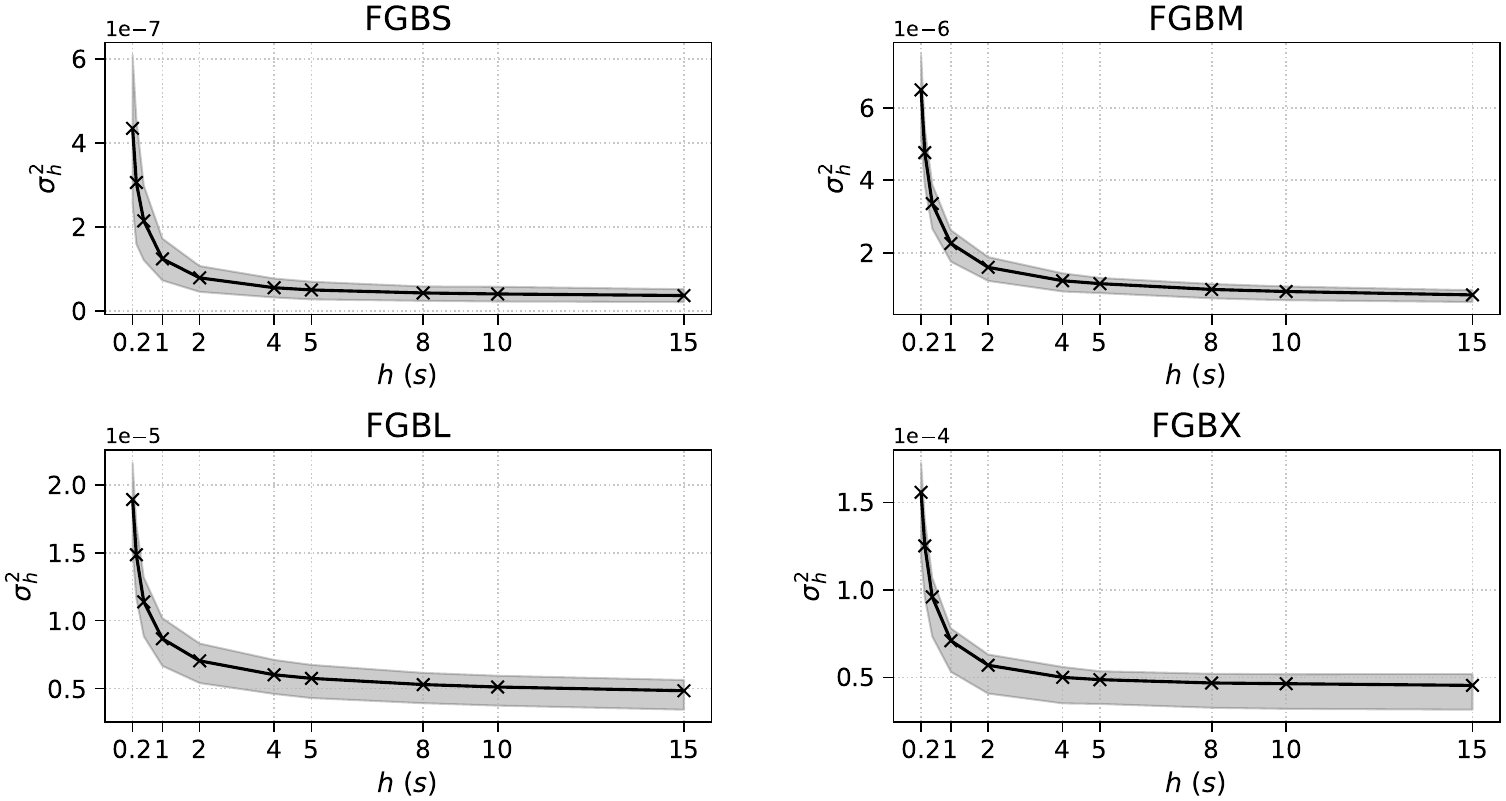}
	\caption{Signature plot.}
	\label{fig:Sign plot normalized}
\end{figure}

Figure \ref{fig:Sign plot normalized} presents the signature plot distribution (mean filled with 5th and 95th percentiles) for different sampling frequencies $h$. This figure illustrates that the signature plot is generally consistent for a given instrument across different days (the 5th to 95th percentiles interval is quite narrow). However, the magnitude of values varies across assets, which is to be expected as different markets exhibit different magnitudes of volatility due to factors such as differences in liquidity, size of orders, and tick size. 

As depicted in the signature plots, the volatility of these four financial instruments can be ordered as follows: FGBX, FGBM, FGBL, and FGBS. This order can be explained by the maturity periods of these instruments. FGBX, representing long-term debt, typically exhibits the highest volatility due to its longer maturity period. The longer the maturity, the more sensitive the instrument is to changes in interest rates, leading to higher price volatility. Next in line is FGBL, which represents medium-term debt and has moderate volatility. It is less sensitive to interest rate changes compared to FGBX due to its shorter maturity period. Then FGBM, another medium-term debt instrument but with a shorter maturity period than FGBL, making it less volatile than FGBL but more volatile than FGBS. Lastly, FGBS, which represents short-term debt, is typically associated with the lowest volatility among the four. Short-term instruments are less sensitive to interest rate changes, leading to lower price volatility.

\begin{figure}[htbp]
	\centering
	\includegraphics[width=0.7\textwidth]{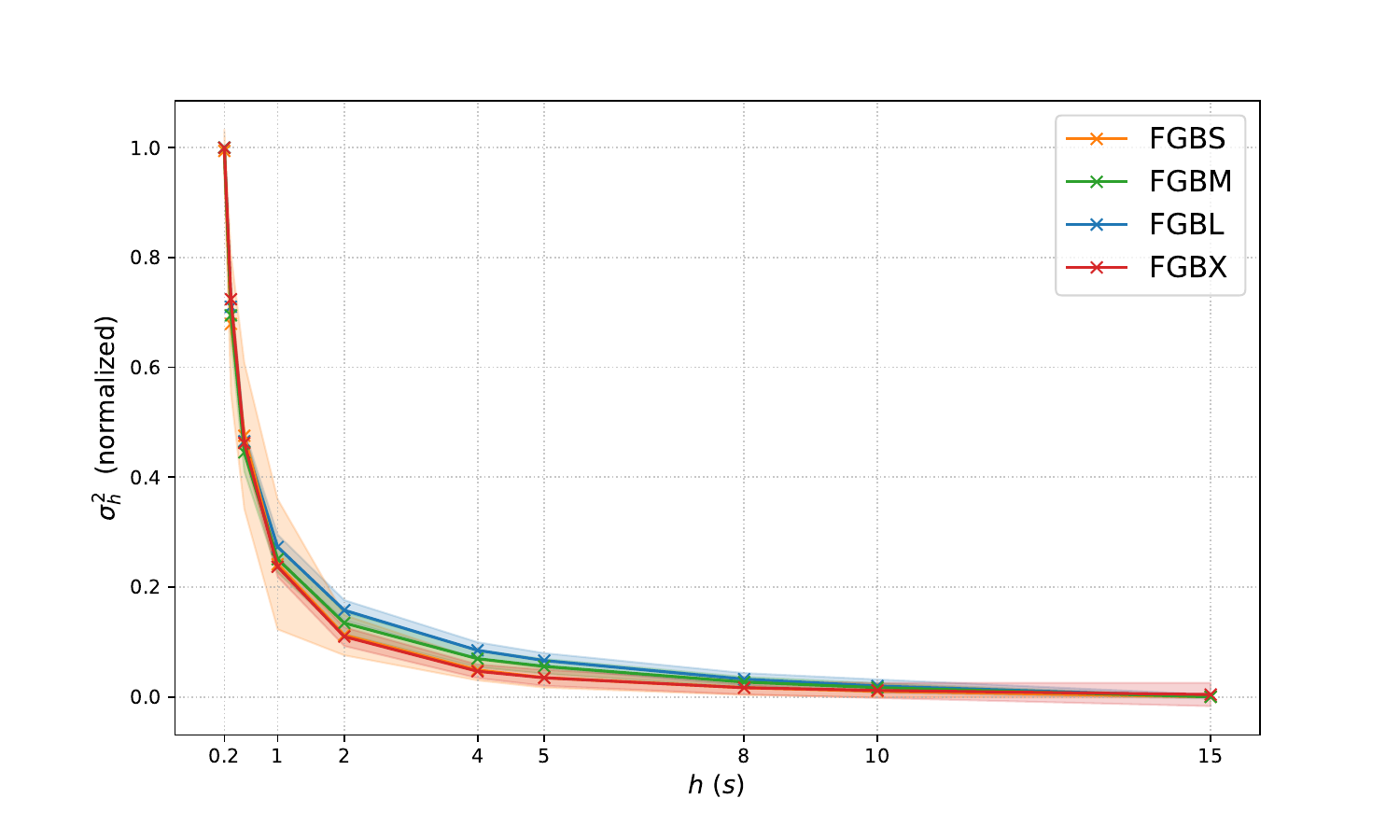}
	\caption{Normalized signature plot.}
	\label{fig:Sign plot aggreg}
\end{figure}

However, despite the differences in volatility among these instruments, a striking similarity is observed when the signature plots are normalized. As shown in Figure \ref{fig:Sign plot aggreg}, which presents the signature plot of all instruments normalized (by dividing by the maximum value), the volatility of the four instruments exhibits a similar decreasing profile.

\subsection{Spread Distributions}

\begin{figure}[htbp]
	\centering
	\includegraphics[width=0.8\textwidth]{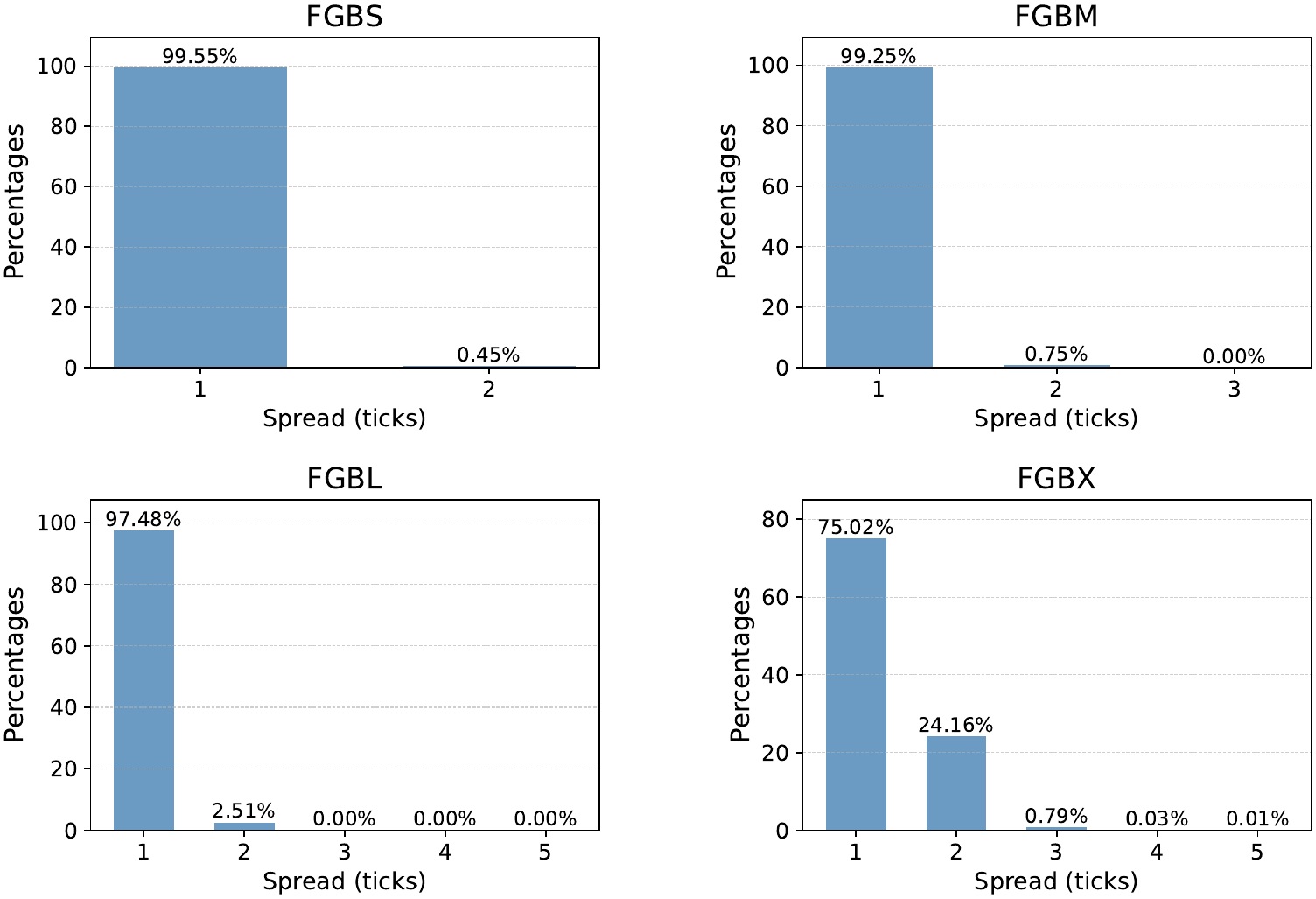}
	\caption{Spread distribution.}
	\label{fig:Tick size distribution}
\end{figure}

The distribution of the bid-ask spread, defined as the difference between the best ask and best bid prices (usually measured in multiples of the price tick size), is a key characteristic of an asset. In the literature, this distribution is often found to decay exponentially \citep{abergel2016limit, bouchaud2018trades}.
However, for assets with large tick sizes, such as the ones subject to this study, the possible spread values are quite limited. Most of the time, the spread is only one tick, and exceptionally large spread values can indicate rare and special events. Figure \ref{fig:Tick size distribution} presents the frequency distribution of spread values.

Particularly for FGBX, the incidence of 2-tick spreads is notably elevated, constituting approximately 24\% of occurrences. This contrasts sharply with other instruments such as FGBL, where it is around 2\%, and less than 1\% for both FGBM and FGBS. One possible explanation for this disparity might be the combination of the long-term maturity of the FGBX asset, which may render it more volatile due to its heightened sensitivity to fluctuations in interest rates, and a relatively low tick size value. This combination of factors likely contributes to the distinct behavior observed in FGBX. Interestingly, the discrepancy in maturity between the FGBM and FGBX instruments (from roughly 5 years to approximately 30 years)  does not align with the variation in their tick sizes (from 0.01 to 0.02).

\begin{figure}[H]
	\centering
	\includegraphics[width=0.82\textwidth]{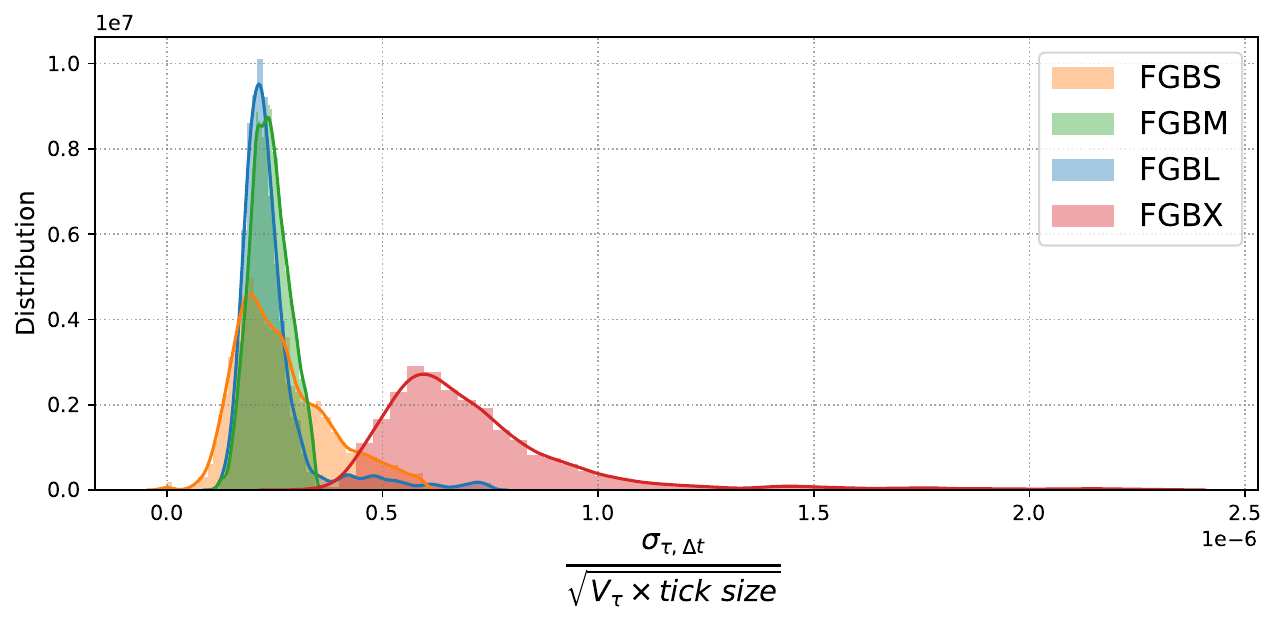}
	\caption{Distribution of  $\frac{\sigma_{\tau, \Delta t}}{\sqrt{V_\tau \times \mathrm{tick~ size}}}$.}
	\label{fig:sigma_vol_ticksize}
\end{figure}

To further investigate this phenomenon, one may consider the ratio \\$\frac{\sigma_{\tau, \Delta t}}{\sqrt{V_\tau \times \mathrm{tick~ size}}}$, where a comparison across various instruments reveals a remarkable similarity, with the exception of FGBX, which exhibits substantially larger values. Figure \ref{fig:sigma_vol_ticksize} provides an illustration of the distribution of this ratio for $\tau = 1h$ and $\Delta t = 1\mathrm{ min}$. The figure clearly emphasizes the particularly high values for the FGBX instrument, thereby underlining the argument that the tick size for this instrument is perhaps particularly underestimated.

\subsection{Spread Duration}

\begin{figure}[H]
	\centering
	\begin{subfigure}[b]{0.75\textwidth}
  \includegraphics[width=\textwidth]{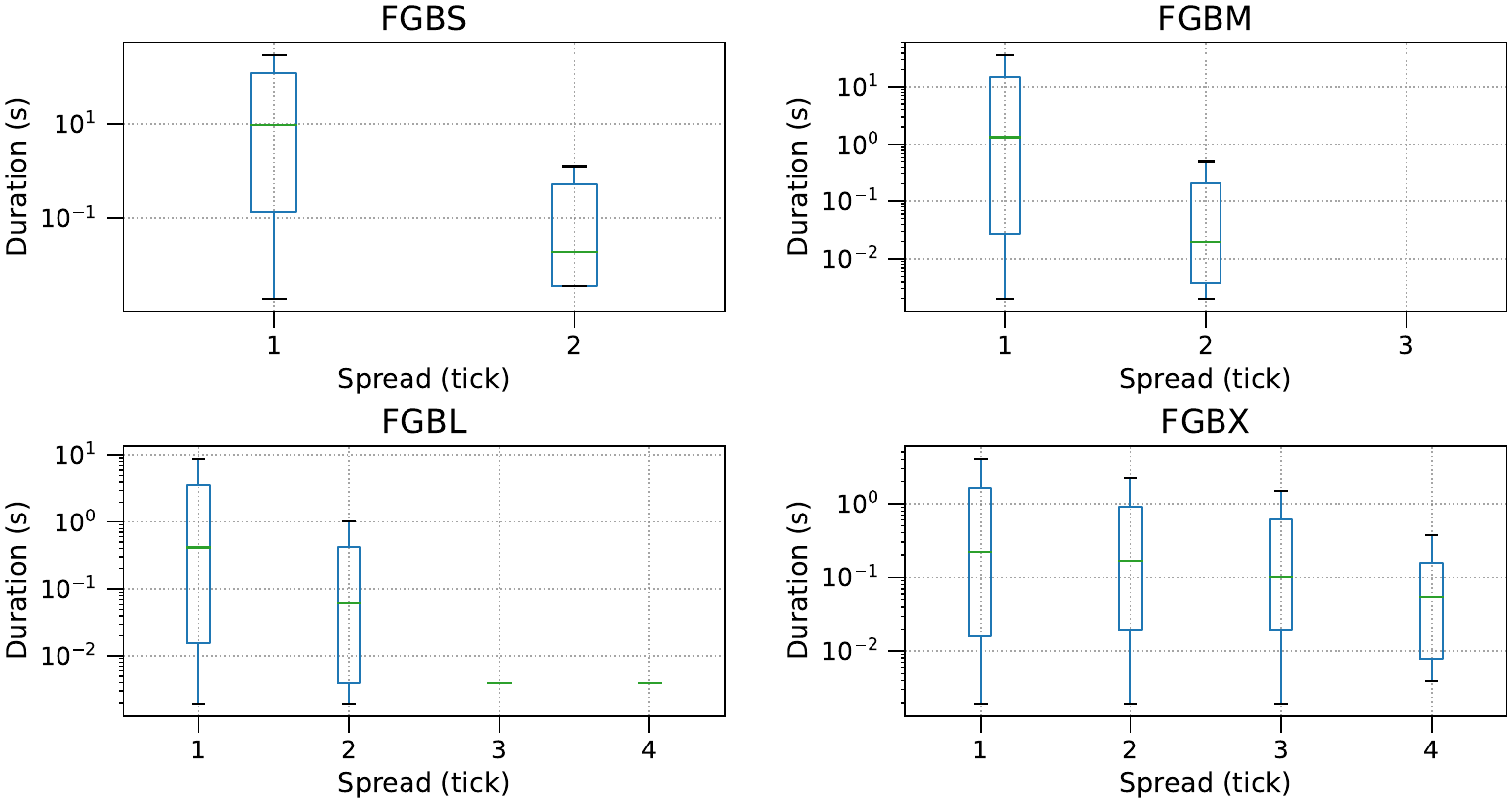}
	\caption{Calendar time.}
	\label{fig:spread duration calendar}
	\end{subfigure}
	\hfill 
	\begin{subfigure}[b]{0.75\textwidth}
	\includegraphics[width=\textwidth]{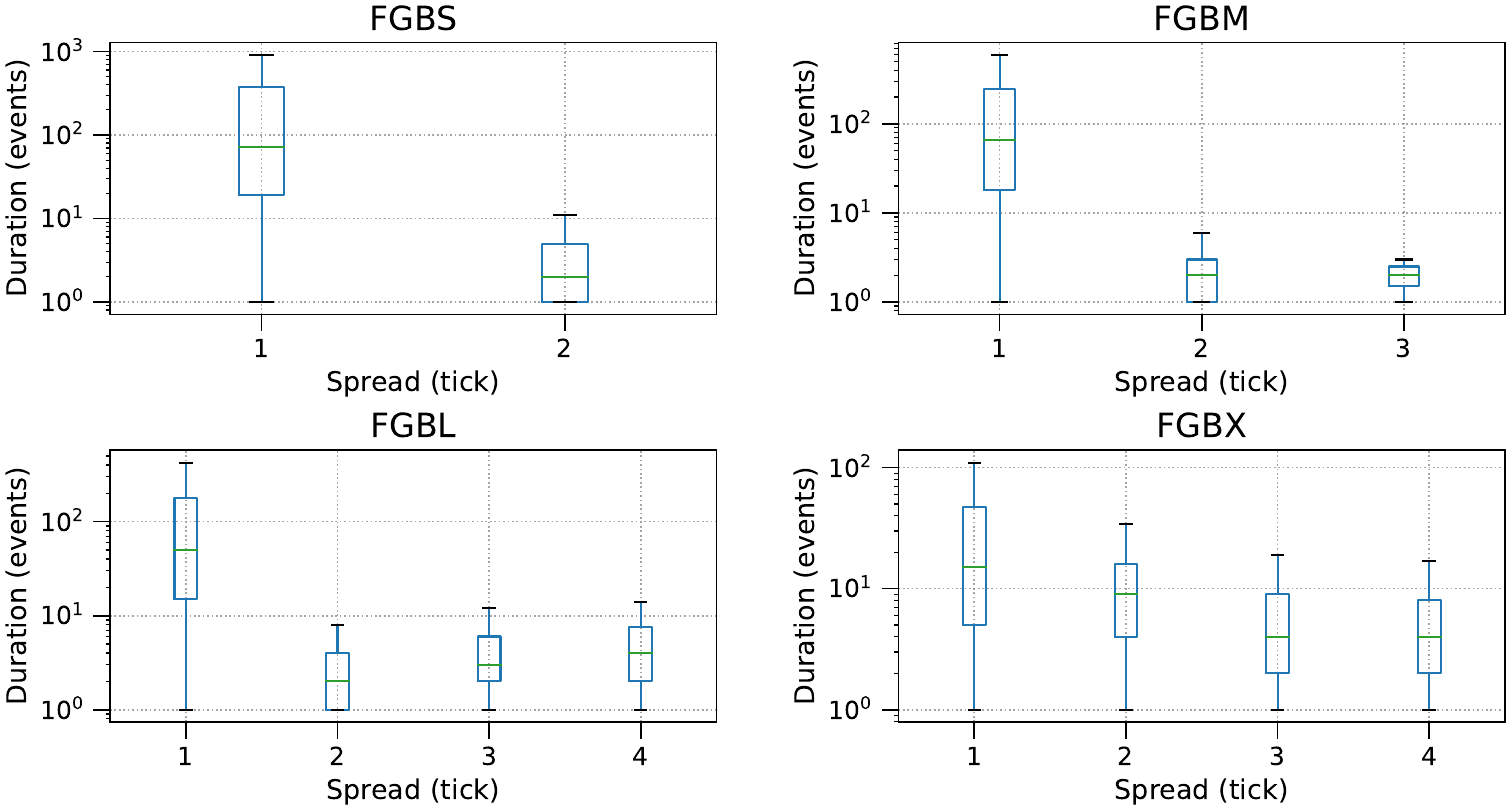}
  	\caption{Event time.}
  	\label{fig:spread duration event}
	\end{subfigure}
	\caption{Spread duration in calendar time (a) and event time (b).}
\end{figure}

Another intriguing characteristic of large tick assets is the notably short duration of large spreads compared to the duration of minimal spreads (1 tick spread). Figures \ref{fig:spread duration calendar} and \ref{fig:spread duration event} illustrate the box plots of duration, in both physical and event times, for different possible spread values. These figures clearly demonstrate how the duration of a spread decreases as its value increases, with larger spread values lasting 10 to 100 times shorter.

This phenomenon can be attributed to the high liquidity in these markets, which encourages market participants to act swiftly when the spread is large. Additionally, market participants are motivated to secure priority in the queue at the new price level by quickly placing orders within the spread. This rush to place orders can significantly reduce the duration of large spreads.

Furthermore, the presence of algorithmic and high-frequency trading can also contribute to this phenomenon. These trading strategies often involve placing and canceling orders rapidly in response to changes in market conditions, which can lead to shorter duration for large spreads.

\subsection{Order Book Shape}

\begin{figure}[htbp]
	\centering
	\includegraphics[width=0.75\textwidth]{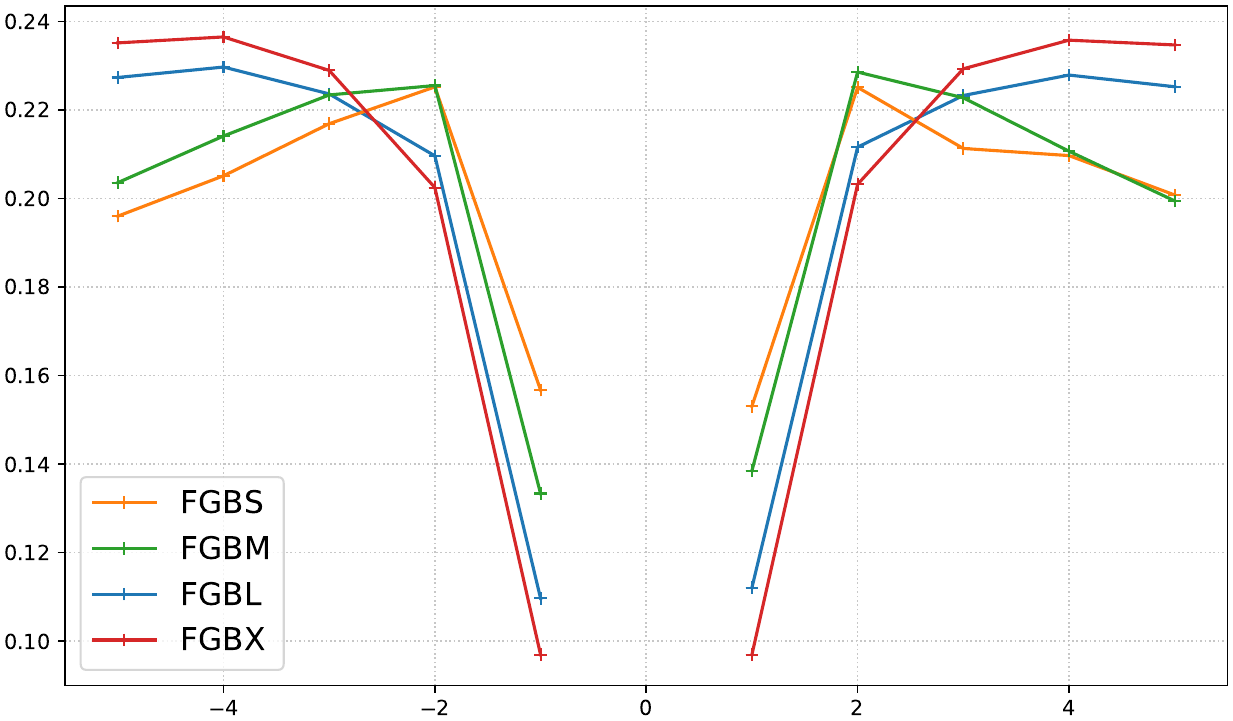}
	\caption{Average quantity offered in the limit order book.}
	\label{fig:ob_shape}
\end{figure}

As illustrated in Figure \ref{fig:ob_shape}, the distribution of liquidity across the various levels of the order book exhibits unique characteristics for each instrument. In the case of FGBL and FGBX, there is a discernible increase in average volume with depth, a pattern aligning with the commonly accepted notion that deeper levels of the order book are often associated with greater standing liquidity \citep{bouchaud2002statistical, abergel2016limit}. This phenomenon may be explained by market participants positioning larger orders at a distance from the best bid and ask prices, in anticipation of more advantageous opportunities.

Contrastingly, with FGBS and FGBM, the average volume demonstrates an initial increase but begins to diminish starting from the third limit. This behavior suggests that the liquidity distribution in the order book for these short-term interest rate products could be shaped by distinct factors, including a perception of reduced risk and the particular trading strategies adopted by market participants. Indeed, these markets are rarely penetrated (as depicted in Figure \ref{fig:Tick size distribution}), which discourages market participants from placing orders very deep in the order book; instead, they exhibit a greater interest in the second level. Moreover, this variance in the configuration of the order book might also mirror the differential behaviors of market makers when dealing with short-term futures as compared to long-term ones.

\section{Conclusion}

In this study, we have embarked on an in-depth exploration of the microstructure of the futures market, with a specific focus on four German bond futures: FGBS, FGBM, FGBL, and FGBX. Our analysis has revealed a rich array of shared characteristics among these instruments, yet each one retains a distinctive character. This distinctiveness is a direct consequence of the intrinsic differences they possess, such as dependence in interest rates, maturity periods, and tick sizes.

Our findings underscore the importance of recognizing and accounting for the unique characteristics of each financial instrument when analyzing market dynamics or developing trading strategies. The stylized facts we have identified, from the distribution of order sizes to the patterns of order flow and interarrival times, are not mere statistical curiosities. They are fundamental aspects of market behavior that have profound implications for market participants and modelers.

In conclusion, our study illuminates the complex dynamics of the German bond futures market and the distinct characteristics of the financial instruments within it. We are confident that these insights will prove useful for those aiming to deepen their understanding of these markets or develop market simulators that are both more accurate and more realistic. It is essential for such simulators to not only validate these stylized facts but also adhere to the range of parameters specific to each instrument. This adherence is a critical factor in determining the realism of the simulators, ensuring they provide a faithful representation of the market dynamics.

\begin{samepage}
\section*{Acknowledgements}

I would like to thank my PhD supervisor, Olivier Guéant, for his guidance and insightful discussions during this project. I am also grateful to my colleagues at BNP Paribas, particularly Guillaume Bioche, for their assistance in interpreting some of the study's results and advising on its refinement.

\end{samepage}

 \bibliographystyle{elsarticle-num-names}
 \bibliography{cas-refs}





\end{document}